\documentclass[fleqn,usenatbib,useAMS]{mnras}

\usepackage[T1]{fontenc}
\usepackage{ae,aecompl}

\usepackage{graphicx}	
\usepackage{amsmath}	
\usepackage{amssymb}	

\usepackage{multicol}
\usepackage{pdfpages}
\usepackage{pdflscape}

\usepackage{newtxtext,newtxmath}

\usepackage{enumerate}
\usepackage{lipsum}
\usepackage{mathrsfs}

\usepackage[figuresright]{rotating}
\usepackage{animate}
\usepackage{lineno}
\usepackage{color}
\usepackage{verbatim}
\usepackage{geometry}
\usepackage{hyperref}

\usepackage{booktabs}
\usepackage{rotating}
\DeclareRobustCommand{\VAN}[3]{#2}
\let\VANthebibliography\thebibliography
\def\thebibliography{\DeclareRobustCommand{\VAN}[3]{##3}\VANthebibliography}

\def\kms{km s$^{-1}$}

\def\cc{{\rm cm^{-3}}}
\def\sc{{\rm cm^{-2}}}

\def\nht{{\rm NH_3}}

\title[Hyperfine Group Ratio (HFGR)]{Hyperfine Group Ratio (HFGR): A Recipe for Deriving Kinetic Temperature from the Ammonia Inversion Lines}

\author[Shen Wang et al.]{
Shen Wang\thanks{Email: shenwang@mail.ustc.edu.cn}$^{1,2}$,
Zhiyuan Ren\thanks{Email: renzy@nao.cas.cn}$^{1*}$,
Di Li\thanks{Email: dili@nao.cas.cn}$^{1,2,3*}$,
Jens Kauffmann,$^{4}$
Qizhou Zhang$^{5}$,
and Hui Shi$^{1,2}$
\\
$^{1}$ National Astronomical Observatories, Chinese Academy of Sciences, A20 Datun Road, Chaoyang District, Beijing 100101, China \\
$^{2}$ University of Chinese Academy of Sciences, Beijing 100049, People’s Republic of China \\
$^{3}$ NAOC-UKZN Computational Astrophysics Centre, University of KwaZulu-Natal, Durban 4000, South Africa \\
$^{4}$ Haystack Observatory, Massachusetts Institute of Technology, 99 Millstone Road, Westford, MA 01886, USA \\
$^{5}$ Center for Astrophysics | Harvard \& Smithsonian, 60 Garden Street, Cambridge, MA 02138, USA
}

\date{Accepted 2020 September 26. Received 2020 September 26; in original form 2020 June 25}

\pubyear{2020}

\begin{document}
\label{firstpage}
\pagerange{\pageref{firstpage}--\pageref{lastpage}}
\maketitle

\begin{abstract}
Although ammonia is a widely used interstellar thermometer, the estimation of its rotational and kinetic temperatures can be affected by the blended Hyperfine Components (HFCs). We developed a new recipe, referred to as the HyperFine Group Ratio (\emph{HFGR}), which utilizes only direct observables, namely the intensity ratios between the grouped HFCs. As tested on the model spectra, the empirical formulae in \emph{HFGR} can derive the rotational temperature ($T_{\rm rot}$) from the HFC group ratios in an unambiguous manner. We compared \emph{HFGR} with two other classical methods, \emph{intensity ratio} and \emph{hyperfine fitting}, based on both simulated spectra and real data. \emph{HFGR} has three major improvements. First, \emph{HFGR} does not require modeling the HFC or fitting the line profiles, thus is more robust against the effect of HFC blending. Second, the simulation-enabled empirical formulae are much faster than fitting the spectra over the parameter space, so the computer time and human time can be both largely saved. Third, the statistical uncertainty of the temperature $\Delta T_{\rm rot}$ as a function of the signal-to-noise ratio (SNR) is a natural product of the \emph{HFGR} recipe. The internal error of \emph{HFGR} is $\Delta T_{\rm rot}\leq0.5$ K over a broad parameter space of rotational temperature (10 to 60 K), line width (0.3 to 4 \kms), and optical depth (0 to 5). When there is a spectral noise, \emph{HFGR} can also maintain a reasonable uncertainty level at $\Delta T_{\rm rot}\leq 1.0$ K (1 $\sigma$) when SNR~$>4$.

\end{abstract}




\section{Introduction}
\label{sec:intro}
Gas temperature is a fundamental parameter of molecular clouds. An accurate temperature measurement is indispensable for studying all the physical and chemical aspects of a cloud. As the first polyatomic molecule discovered in interstellar medium \citep{cheung68}, Ammonia ($\nht$) is considered to be an ideal temperature tracer for the dense molecular gas \citep{ho83,li03,mangum15}. The rotational levels $(J,K)$ of the para-$\nht$ have largely different excitation energies, but the inversion transition of each $(J,K)$ level is distributed in a relatively small frequency range of 23-25 GHz. In the meantime, the $\nht$ inversion lines are split into hyperfine components (HFCs) due to the varied alignment between the nitrogen and hydrogen nuclei \citep[][also see Figure~\ref{fig:trans_level} ]{ho83,li03,mangum15}. The HFCs can be used to estimate the optical depth so that the temperature measurement become more accurate \citep{ho83}. Due to these attributes, $\nht$ is an invaluable tool of probing the physical conditions in molecular clouds.  Over three decades, extensive $\nht$ surveys have been carried out to map the giant molecular clouds \citep{purcell12, seo15, friesen17, hogge18} and the individual dense molecular cores at different evolutionary stages \citep{myers83a, jijina99, wienen18, svoboda16}. These surveys demonstrate the reliability of the $\nht$ inversion transitions in tracing the spatial and temperature distributions of the dense molecular gas in different spatial scales (from several parsec to $10^3$ AU) and in a broad temperature range (10 to 50 K).

There are two classic methods for the temperature calculation: (1) \emph{Intensity ratio}, which starts from the observed spectra, using the integrated intensity ratios between high- and low-excitation lines to derive the temperature based on Boltzmann distribuiton \citep[e.g.][]{ho83,busquet09,ragan11,williams18,dhabal19}; (2) Hyperfine fitting (denoted as \emph{HF fitting} hereafter), which uses radiative transfer functions to generate the model spectra, and the observed spectra can be fitted by adjusting the temperature and other parameters \citep[e.g.][]{rosolowsky08,estalella17,keown17,camacho20}. One can perform the spectral fitting using the Python Package Pyspeckit \citep{ginsburg2011}, or the compiled package \emph{HfS} \citep{estalella17}. \emph{HfS} is featured by an easy operation of separating different velocity components and estimating the parameters of each component respectively.

\emph{Intensity ratio} and \emph{HF fitting} are both widely used for temperature measurement in a variety of molecular clouds. But they still have some major aspects to be improved. \emph{Intensity ratio} method involves an approximation that each HFC group has one single average intensity and optical depth, which may cause potential uncertainty (described in Section \ref{sec:compare}). \emph{HF fitting} requires to traverse over a broad parameter space to find the optimized values, which would be time-consuming, in particular when data size is large.

We tried to improve the accuracy and efficiency in calculation by building a recipe, which is based on more direct connection between the physical parameters and observed line intensities. The connections between the spectral line profiles and physical parameters include two major aspects:  

(1) the intensity ratio between high- and low-excitation lines increases with $T_{\rm rot}$; 

(2) the intensity ratio between optically thin (satellites) and thick (main) HFCs increases with the optical depth. 

In the current work, we try to more directly use these two relations instead of rely on additional assumptions about the optical depths.  

In modeling the spectra, we found that $T_{\rm rot}$ can sensitively determine the HFC intensities and its effect is not degenerated with the optical depth, so that the recipe can be expressed in a group of empirical formulae, which are as simple as \emph{Intensity ratio}, but more stable over a broad parameter space. The recipe is named as the \emph{Hyperfine Group Ratio (HFGR)}. It is described in details in the following sections. The physical background of the $\nht$ inversion lines is described in Section \ref{sec:bg}. The spectra modeling is introduced in \ref{sec:spec_model}. The empirical formulae of $T_{\rm rot}$ are derived from the spectra in Section \ref{sec:r_sm} to \ref{sec:hfgr_use} and its intrinsic accuracy is examined in Section \ref{sec:hfgr_accuracy} to \ref{sec:hfgr_low_trot}. A comparison between \emph{HFGR} and other two methods for their accuracy and efficiency are presented in \ref{sec:compare}. And a further comparison based on the observed data is presented in \ref{sec:obs_data}. A summary is given in Section \ref{sec:summary}. 

\section{Background of the $\nht$ Temperature Calculation}\label{sec:bg}
\subsection{Physical Basis of the $\nht$ Inversion Transitions}
$\nht$ is a symmetric top molecule and has relatively simple rotational energy level structures. The physical basis of the population and line transitions are revealed and discussed in a series of papers \citep{kukolich67,rydbeck77,ho83,stahler05,mangum15}.  The major physical basis is that the $\nht$ rotational energy levels are characterized by the quantum numbers of total angular momentum $J$ and its projection along the molecular axis $K$. The para-$\nht$ has the two lowest metastable levels with quantum numbers of $(J,K)=(1,1)$ and (2,2), while the $(J,K)=(3n, 3n)~(n=0,1,2,3...)$ levels are populated by ortho-$\nht$. The $(1,1)$ and $(2,2)$ levels have an energy difference of $\Delta E/k=41.5$ K. The populations on the (1,1) and (2,2) levels mainly rely on collisional excitation. In typical cold dense clouds with number density of $n(\rm{H_2})\sim 10^4~\rm{cm^{-3}}$ and $T_{\rm gas}\leq15$ K \citep[e.g.][]{friesen17}, the $\nht$ molecules would be mainly populated on the (1,1) and (2,2) levels.

\begin{figure*}
\centering
\includegraphics[angle=0,width=0.9\textwidth]{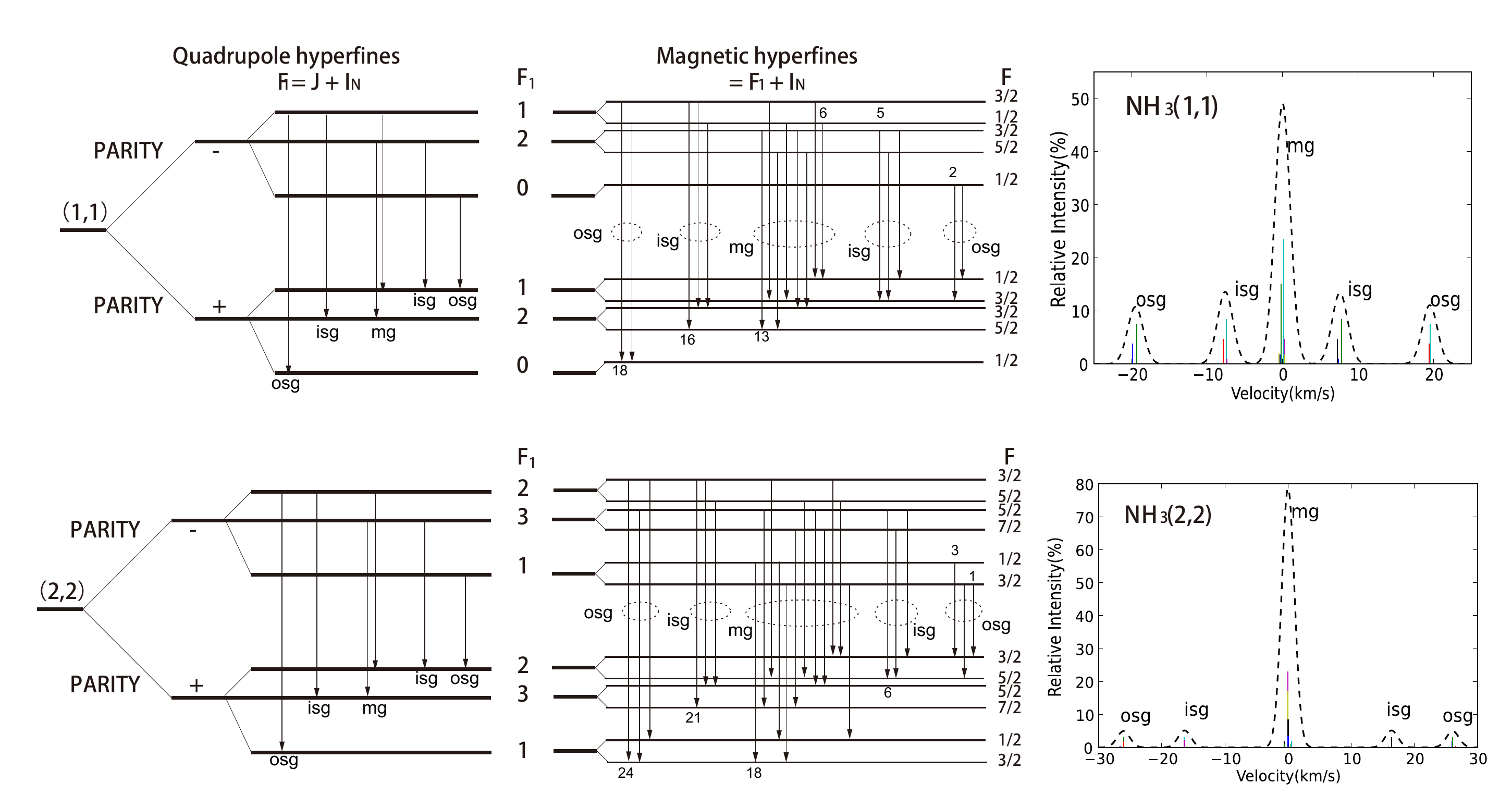}   \\ 
\caption{\small The energy levels, hyperfine splitting, and the transitions of the $\nht$ (1,1) and (2,2) levels (left and middle panels), and the locations of the hyperfine groups locations in the spectral profiles (right panels). }
\label{fig:trans_level}
\end{figure*}

The (1,1) and (2,2) levels both have transitions between different parities of the nitrogen wave function over the plane of three hydrogen atoms. The transitions are further split into a series of HFCs, which are specified in \citet{rydbeck77} and shown in Figure~\ref{fig:trans_level}. There are 18 HFCs in the $J_K^P=(1_1^--1_1^+)$ transition and 24 HFCs in the $J_K^P=(2_2^--2_2^+)$. The right panels in Figure~\ref{fig:trans_level} show the synthetic spectra for the two transitions. For both the $(1_1^--1_1^+)$ and $(2_2^--2_2^+)$ lines, the HFCs can be divided into the main group (mg), inner satellite group (isg) and outer satellite group (osg), as labeled in Figure~\ref{fig:trans_level}. And these transitions have similar frequencies thus can be usually observed at the same time. 

The intrinsic strengths of the HFC groups are measured from the laboratory spectra assuming all the transitions to have the same excitation temperature, total gas number density and $\nht$ column density. We adopt the normalized line strengths listed in the review of \citet[][Table 19 and Table 20 therein]{mangum15}. The information of 18 HFCs in $(1_1^--1_1^+)$ and 24 HFCs in $(2_2^--2_2^+)$ are presented in Table \ref{tbl:hf_pars} and \ref{tbl:line_pars}.

\subsection{The Previous Methods: Aspects to be Improved} 
\begin{figure*}
\centering
\includegraphics[angle=0, width=0.7\textwidth]{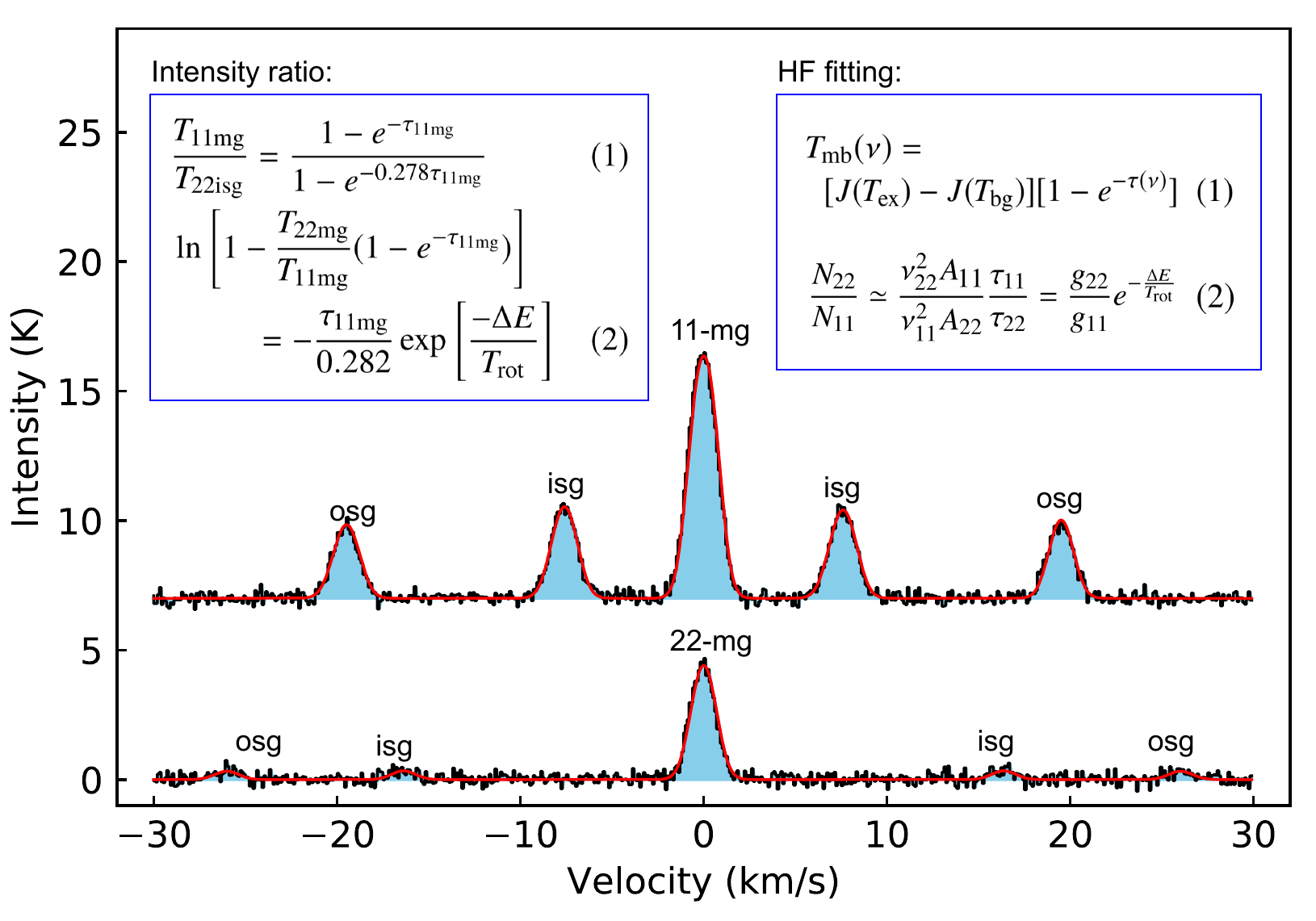}   \\  
\caption{\small An example of $(1_1^--1_1^+)$ and $(2_2^--2_2^+)$ spectra, with HF groups are labeled on each HF group. The red solid lines represent the best-fit spectra from the method of \emph{HF fitting}. The major equations in the two methods are also presented on the figure, including the equations used to calculate $T_{\rm rot}$ from the HF groups in \emph{Intensity ratio} and the equations to produce the model spectra in \emph{HF fitting}.}
\label{fig:spec_hf}
\end{figure*}

An example of the $\nht$ $(1_1^--1_1^+)$ and $(2_2^--2_2^+)$ spectra and the major equations used by \emph{Intensity ratio} and \emph{HF fitting} methods are shown in Figure \ref{fig:spec_hf}. The \emph{Intensity ratio} method uses the intensities of the HFCs to derive $T_{\rm rot}$ assuming that the two levels obey the Boltzmann distribution \citep{ho83,mangum92,ragan11}, that is
\begin{equation}    
\begin{aligned}
T_{\rm rot} & = & \quad \\
\quad & -\Delta E/k \div \ln\left[-\frac{0.282}{\tau(1,1,{\rm mg})}\ln[1-\frac{T_B(2,2,{\rm mg})}{T_B(1,1,{\rm mg})}(1-e^{-\tau(1,1,{\rm mg})})] \right]. & \quad \\
\end{aligned}
\end{equation}

The $(1_1^--1_1^+)$ transition could be moderately optically thick. Its optical depth can be estimated from the ratio between the main and satellite groups of HFCs in \citep[][also see Figure~\ref{fig:spec_hf}]{ho83}
\begin{equation}    
\begin{aligned}
\frac{T_B(1,1,{\rm mg})}{T_B(1,1,{\rm isg})} & = \frac{1-e^{-\tau(1,1,{\rm mg})}}{1-e^{-\tau(1,1,{\rm isg})}} & \quad \\
\quad & = \frac{1-e^{-\tau(1,1,{\rm mg})}}{1-e^{-a\tau(1,1,{\rm mg})}} = \frac{1-e^{-\tau(1,1,{\rm mg})}}{1-e^{-0.278\tau(1,1,{\rm mg})}}, & \quad \\
\end{aligned}
\end{equation}
where $a=0.278$ is the intensity ratio between $\nht$ $(1_1^--1_1^+)$ main and inner satellite groups. $T_B(1,1,{\rm mg})$ and $T_B(1,1,{\rm isg})$ are the observed brightness temperatures of the two groups, respectively. 

\begin{figure*}
\centering
\includegraphics[angle=0,width=0.9\textwidth]{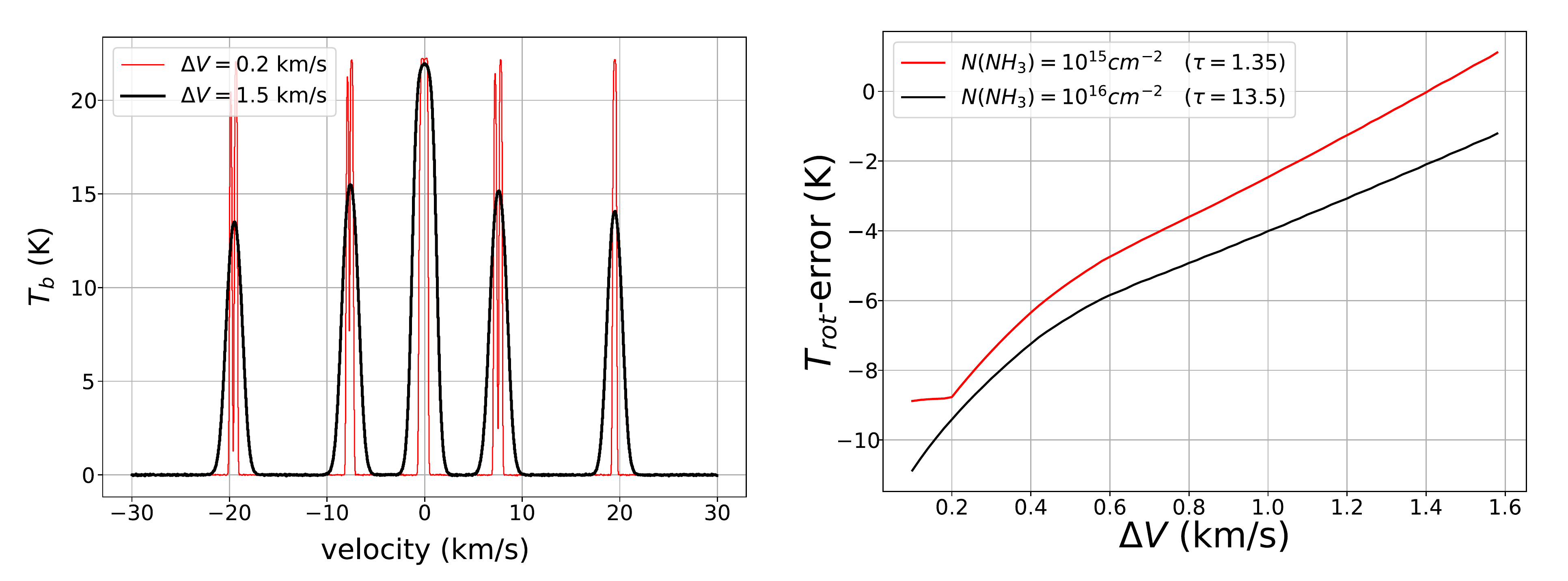} \\   
\caption{\small The $T_{\rm rot}$ deviation as a function of the line width $\Delta V$. {\bf (a)} two model spectra to elucidate how $\Delta V$ can alter the line profile and HF-group intensities. {\bf (b)} The $T_{\rm rot}$ deviation as a function of $\Delta V$ in \emph{Intensity ratio} method. The real temperature is $T_{\rm rot}=20$ K. The deviation curve is calculated at the two values of the optical depths. }
\label{fig:spec_dv}
\end{figure*}

Equation (2) assumes that each HFC group has a unique optical depth $\tau_{\rm group}$. This is an approximation since each group actually contains several internal HFCs with slightly different frequencies (Table~\ref{tbl:hf_pars} and Figure~\ref{fig:trans_level} right panels), corresponding to an average velocity difference of $\overline{\Delta V}_{\rm HFC}\simeq 0.3$ \kms. If the line width is relatively large ($\Delta V>\overline{\Delta V}_{\rm HFC}$), the HFCs would be overlapped and the group could be regarded as an integrate spectral feature. In contrast, if $\Delta V<\overline{\Delta V}_{\rm HFC}$, the internal HFCs would be further separated and have individual $\tau$ values, which are not necessarily equal. In this case, it would be less accurate to assume each HFC group to have a single value of $\tau_{\rm group}$. The two cases, namely separated and overlapped HFCs, are presented in Figure~\ref{fig:spec_dv}a. 

For \emph{HF fitting}, we estimated its $T_{\rm rot}$ deviation due to $\Delta V$. The $\nht$ $(1_1^--1_1^+)$ and $(2_2^--2_2^+)$ model spectra can be generated using the radiative transfer functions. The physical parameters to determine the spectra include optical depth $\tau_0$, rotational temperature $T_{\rm rot}$, and the intrinsic line width $\Delta V$. The optical depth as a function of the frequency is assumed to have a Gaussian profile for each HFC, that is
\begin{equation}     
\tau(\nu)=\tau_0 \sum_i s_i \exp \left[-\left(\frac{\nu-\nu_i-\nu_0}{2 \sigma_\nu^2}\right)^2\right], 
\end{equation}
wherein $\tau_0$ is the central optical depth of the $(1_1^--1_1^+)$ transition, $\nu_0$ is the observed central frequency of the mg component, $s_i$ is the relative intensity of each HFC, and $\nu_i$ is the frequency shift of each HFC relative to $\nu_0$. The values of $\nu_i$ and $s_i$ are adopted from \citet{kukolich67} and are shown in Table 1. The spectral frequency is related to the radial velocity as $(\nu_0-\nu)/\nu_0$=$(V-V_0)/c$. And hence is the frequency and velocity dispersions, $\sigma_\nu/\nu_0$=$-\sigma_V/c$. And $\sigma_V$ is related to the full-width half-maximum (FWHM) line width $\Delta V$ as $\sigma_v = \Delta V /\sqrt{8 \ln 2}$. Using the Planck-corrected brightness temperature
\begin{equation}     
J(T)=\frac{h\nu_{ul}/k}{\exp(h\nu_{ul}/kT)-1},
\end{equation}
the $(1_1^--1_1^+)$ and $(2_2^--2_2^+)$ spectra can be modeled using the radiative transfer function as:
\begin{equation}     
T_{\rm mb}(\nu)=\eta_f[J(T_{\rm ex})-J(T_{\rm bg})][1-e^{-\tau(\nu)}],
\end{equation}
Wherein $\tau(\nu)$ is the optical depth as a function of frequency $\nu$, $T_{\rm ex}$ is the excitation temperature, $T_{\rm bg}=2.73$ K is the cosmic background temperature, $\eta_f$ is the beam filling factor and is set to be $\eta_f=1$ in modeling the line profile. 

The output $T_{\rm rot}$ is calculated from the model spectra using Equation (1) and (2). Figure~\ref{fig:spec_dv}b exhibits the variation of output $T_{\rm rot}$ as a function of $\Delta V$. As expected, when $\Delta V$ is relatively small, the output $T_{\rm rot}$ has a large error of $\Delta T_{\rm rot}=8$ to 10 K. As shown in Figure~\ref{fig:spec_dv}b, $\Delta T_{\rm rot}$ is also affected by $\tau$. When $\tau$ increases by a factor of 10, the derived $T_{\rm rot}$ would have a variation of $\Delta T_{\rm rot}=2$ K. The variation of $\Delta T_{\rm rot}$ suggests the requirement to improve \emph{Intensity ratio} method, in particular when $\Delta V$ is small.

The \emph{HF fitting} method, in comparison, requires a traversal over the broad parameter space of ($\Delta V$, $T_{\rm rot}$, $N(\nht)$) to look for the best-fit spectra. The complexity of this algorithm should be proportional to the parameter space, which is estimated to be $O[n({\rm channels)} \times n(\Delta V) \times n(T_{\rm rot}) \times n(N(\nht))]$, wherein $n({\rm channels})$ is the number of channels in the spectrum. For other parameters, i.e., $\Delta V$, $T_{\rm rot}$, and $N(\nht)$. $n(X)$ represents the number of data points to be sampled over its suspected range. One can attempt to reduce the calculation by carefully selecting the initial values of the parameters and using Monte Carlo sampling to more quickly approach the optimized values, as adopted by \citet{estalella17}. But the overall complexity of $O(n(X))$ is unlikely to be largely reduced.

In this work, we try to improve the $T_{\rm rot}$ calculation by utilizing the advantages of the two methods. Following the simplicity of \emph{Intensity ratio}, we also adopt the strategy of using the HF groups to derive $T_{\rm rot}$, but we do not assume an average optical depth for each HF group. Instead, we considered the more direct connections between $T_{\rm rot}$ and the HFC intensities as mentioned in Section \ref{sec:intro}. For the $(1,1)$ and $(2,2)$ levels, they turn out to be 

(1) the intensity ratio of $T_{\rm mb}(2,2)/T_{\rm mb}(1,1)$ lines increases with $T_{\rm rot}$; 

(2) the intensity ratio of $T_{\rm mb}(1,1,{\rm sg})/T_{\rm mb}(1,1,{\rm mg})$ increases with the optical depth or column density $N(\nht)$. 

Like in \emph{HF fitting}, one needs to adjust the physical parameters in the model spectra to fit these two relations. But once they can be described in empirical formulae with acceptable accuracy, the formulae would be adopted to be an independent method to derive $T_{\rm rot}$, and the spectral fitting over the parameter space is no longer needed.  
 
\section{Recipe for $\nht$ Rotational Temperature}
\subsection{Modeling the $\nht$ inversion spectra}\label{sec:spec_model}
There are three major steps to build the new recipe of estimating $T_{\rm rot}$, including: 

(1) generating the model spectra using radiation transfer functions based on the input parameters, namely Equation (3) to (5); 

(2) sampling the relation between $T_{\rm rot}$ and the HFC intensities from the model spectra to build the empirical formulae; 

(3) evaluating the $T_{\rm rot}$ uncertainty in the empirical formulae and its variation over the parameter space. 

In the condition of Local Thermal Equilibrium (LTE), the (1,1) and (2,2) levels of the obey the Boltzmann distribution. Their column-density ratio is thus
\begin{equation} 
\frac{N(2,2)}{N(1,1)}=\frac{g_{22}}{g_{11}}{\exp\left[-\frac{\Delta E}{k T_{\rm rot}}\right]},
\end{equation}
wherein $N(1,1)$ and $N(2,2)$ are the total column densities of the $(1,1)$ and $(2,2)$ levels, respectively. $\Delta E/k=(E_{22}-E_{11})/k=40.99~\rm{K}$ is the energy difference; the statistic weight ratio is $g_{11}/g_{22}=3/5$ \citep{pickett98}.  The accuracy of Equation (6) requires the two levels to have thermalized population. As calculated from the collisional excitations \citep{sw85, maret09, shirley15}, the thermalization of the metastable levels mainly depend on the total gas number density. For example, as shown in \citet{maret09}, when the density is low ($n({\rm H_2})\leq 10^4\,\cc$), the two transitions have a difference of $T_{\rm ex,11}-T_{\rm ex,22}>10$ K. It declines to $T_{\rm ex,11}-T_{\rm ex,22}=0$ K when $n({\rm H_2})\geq 5\times10^5\,\cc$. \citet{friesen17} performed a similar calculation by directly solving the equations of statistical equilibrium using RADEX \citep{vdtak07}. In a typical physical condition of $n=10^4\,\cc$, $T_{\rm kin}=15$ K, and $N({\rm p-}\nht)=10^{14}\,\sc$, the temperatures are solved to be $T_{\rm ex}(1,1)=8.5$ K and $T_{\rm ex}(2,2)=6.9$ K. Based on these results, we suggests that Equation (6) could be a reasonable approximation when $n({\rm H_2})\geq 5\times10^4$ cm$^{-3}$. At lower densities, the $(J,K)$ levels are more deviated from LTE and it would become physically unfeasible to derive one single $T_{\rm rot}$.


On the other hand, for each metastable level, the column density is related to the total optical depth as \citep{rosolowsky08}:
\begin{equation}  
\begin{aligned}
N(i,~i) & = \frac{8\pi \nu^2_0}{c^2}\frac{g_l}{g_u}\frac{1}{A_i}\left[1-\exp\left(\frac{h\nu_{\rm i}}{kT_{\rm ex}}\right)\right]^{-1}\int \tau_\nu(i,~i) ~ {\rm d}v \\
\quad & \simeq \frac{8\pi \nu^2_0}{c^2}\frac{g_l}{g_u}\frac{1}{A_i}\left[1-\exp\left(\frac{h\nu_{\rm i}}{kT_{\rm ex}}\right)\right]^{-1} \sqrt{2\pi} \sigma_\nu \tau_0(i,i),
\end{aligned}
\end{equation}   
wherein $i=1,2$ and the second term on the right side is obtained by using the $\tau(\nu)$ expression in Equation (3). 

 
An example of $\nht$(1,1) and (2,2) model spectra is presented in Figure~\ref{fig:spec_hf}. The input parameters are $T_{\rm rot}=T_{\rm ex}=28$ K, $\Delta V=1.5$ \kms, $\tau_{0,11}=2.0$, and an rms noise of 0.1 K. Using Equation (6) and (7), one can derive $\tau_{0,22}=0.44$. For the (2,2) line, the satellite components appear to be much weaker compare to that in the (1,1) spectrum. 


\subsection{Relation between $T_{\rm rot}$ and HFC intensities}\label{sec:r_sm}
Using Equation (5), the total optical depth can be manually connected to the observed line intensity as:
\begin{equation} 
\int \tau_\nu {\rm d}\nu\equiv\frac{\int \tau_\nu {\rm d}\nu}{\int {1-e^{-\tau_\nu}}{\rm d}\nu }\frac{\int T_{\rm mb}(\nu) {\rm d}\nu}{J(T_{\rm ex})-J(T_{\rm bg})}. 
\end{equation}

Using this form, the ratio between the (1,1) and (2,2) column densities can be written as
\begin{equation} 
\begin{aligned}
& \frac{N(1,1)}{N(2,2)} = \\
& \quad \left[\frac{\nu_{11}^2}{\nu_{22}^2}\frac{A_{22}}{A_{11}}\frac{1-\exp\left({\frac{h\nu_{22}}{k T_{\rm ex,22}}}\right)}{1-\exp\left({\frac{h\nu_{11}}{kT_{\rm ex,11}}}\right)}\frac{\frac{\int \tau_\nu(1,1) {\rm d}\nu}{\int {1-\exp[{-\tau_\nu(1,1)}]}{\rm d}\nu}}{\frac{\int \tau_\nu(2,2) {\rm d}\nu}{\int {1-\exp[{-\tau_\nu(2,2)}]}{\rm d}\nu}}\right] \frac{\frac{\int T_{\rm mb}(1,1) {\rm d}\nu}{J(T_{\rm ex,11})-J(T_{\rm bg})}}{\frac{\int T_{\rm mb}(2,2) {\rm d}\nu}{J(T_{\rm ex,22})-J(T_{\rm bg})}} \\
\end{aligned}
\end{equation}

Combining Equation (9) and (6), we have
\begin{equation} 
\begin{aligned}
& \exp\left[\frac{\Delta E}{k T_{\rm rot}}\right] = \\
& \quad C_{\rm ex}\left[\frac{g(2,2)}{g(1,1)}\frac{\nu_{11}A_{22}}{\nu_{22}A_{11}}\frac{\frac{\int \tau_\nu(1,1) {\rm d}\nu}{\int {1-\exp[-\tau_\nu(1,1)]}{\rm d}\nu}}{\frac{\int \tau_\nu(2,2) {\rm d}\nu}{\int {1-\exp[-\tau_\nu(2,2)]}{\rm d}\nu }}\right]\frac{\int T_{\rm mb}(1,1) {\rm d}\nu}{\int T_{\rm mb}(2,2) {\rm d}\nu} \\
\end{aligned}
\end{equation}
wherein the quantities depending on $T_{\rm ex}$ are combined into one factor of $C_{\rm ex}=[1-J(T_{\rm bg})/J(T_{\rm ex,22})]/[1-J(T_{\rm bg})/J(T_{\rm ex,11})]$. We further reduced Equation (10) by assuming an LTE condition so that $C_{\rm ex}=1$. We  estimated the $C_{\rm ex}$ variation with the difference between $T_{\rm ex,11}$ and $T_{\rm ex,22}$. Also adopting the $T_{\rm ex}$ variation scale in \citet{friesen17}, namely $T_{\rm ex}(2,2)-T_{\rm ex}(1,1)=\pm 2$ K, we can derive $C_{\rm ex}=0.9$ to 1.1. That means if the $(1,1)$ and $(2,2)$ excitations are not largely deviated from LTE, we can still have the approximation of $C_{\rm ex}\simeq1.0$.

Equation (10) can be further reduced by defining a correction factor $C_f$ as
\begin{equation} 
C_f=\frac{g_{22}}{g_{11}}\frac{\nu_{11}A_{22}}{\nu_{22}A_{11}} \left[\frac{\int \tau_\nu(1,1) {\rm d}\nu}{\int \tau_\nu(2,2) {\rm d}\nu} \frac{\int {1-e^{-\tau_\nu(2,2)}}{\rm d}\nu}{\int {1-e^{-\tau_\nu(1,1)}}{\rm d}\nu}\right].
\end{equation}

Using $C_f$, Equation (10) becomes
\begin{equation} 
\exp\left[\frac{\Delta E}{T_{\rm rot} k}\right]=C_f\times \frac{\int T_{\rm mb}(1,1) {\rm d}\nu}{\int T_{\rm mb}(2,2) {\rm d}\nu},
\end{equation}
Equation (12) is then transformed into an expression of $T_{\rm rot}$,
\begin{equation} 
T_{\rm rot}=\frac{\Delta E /k}{\ln \left[C_f\times\frac{\int T_{\rm mb}(1,1) {\rm d}\nu}{\int T_{\rm mb}(2,2) {\rm d}\nu}\right]}.
\end{equation}

Now the key step is to express $C_f$ using the observed quantities. {\color{black} Since a main purpose of this work is to circumvent the uncertainty due to the HFC-blending, we consider using the intensity ratios among the HFC groups (mg, isg, osg).} Since $C_f$ is related to the optical depth, a natural option is to consider the intensity ratio between mg and sg which is also proportional to the optical depth, that is
\begin{equation}  
R_{\rm sm}=\frac{\int T_{\rm mb}^{\rm isg+osg} {\rm d}\nu}{\int T_{\rm mb}^{\rm mg} {\rm d}\nu} \bigg|_{(1,1)}.
\end{equation}

Theoretically, the HFC groups in the $(2_2^--2_2^+)$ transition can also estimate $C_f$ as shown in Equation (11). They are not adopted mainly because the satellite groups are much weaker than in $(1_1^--1_1^+)$ (Figure \ref{fig:trans_level} and \ref{fig:spec_hf}). 

The relation between $C_f$ and $R_{\rm sm}$ can be numerically sampled from the model spectra over the $\tau_0$ range. In each sampling over the $\tau_0$ range, the other two parameters, $\Delta V$ and $T_{\rm rot}$ are set to be constants. Then a number of samplings are carried out to obtain the $C_f(R_{\rm sm})$ relation at different $\Delta V$ and $T_{\rm rot}$ values. 

\begin{figure*}
\centering
\includegraphics[angle=0,width=0.9\textwidth]{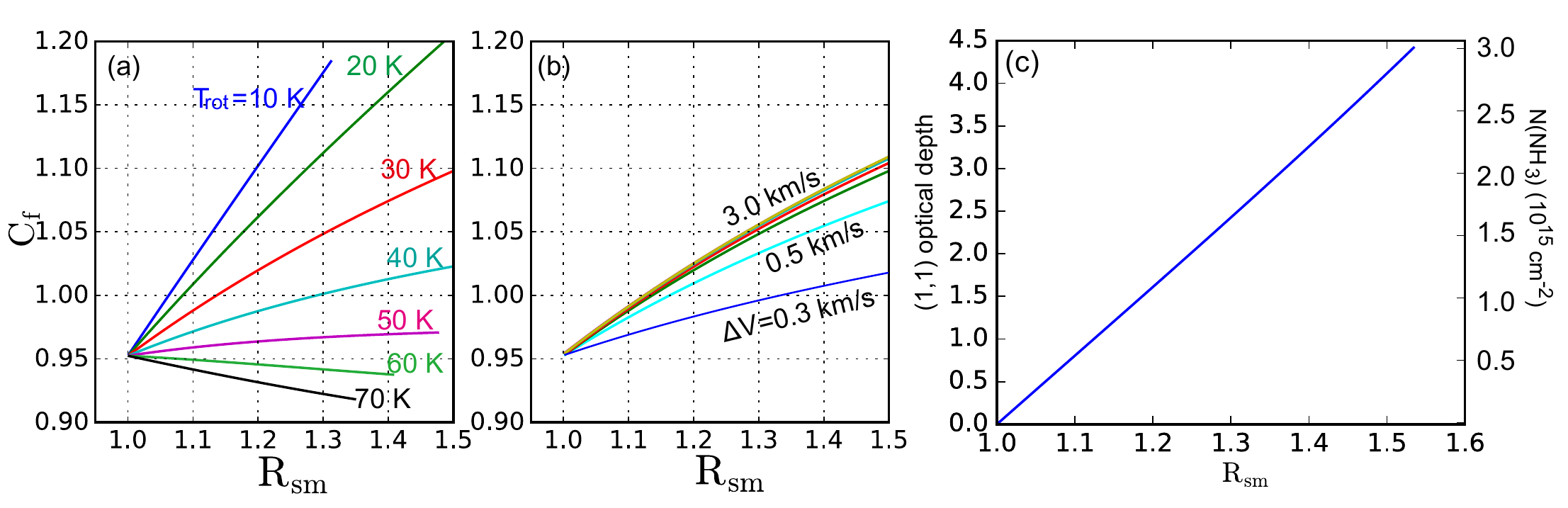}    \\  
\caption{\small {\bf (a)} The correction factor $C_f$ as a function of $R_{\rm sm}$ at different $T_{\rm rot}$ values, with the line width fixed to be $\Delta V=1.0$ \kms and $\tau=2.0$. {\bf (b)} The $C_f$-$R_{\rm sm}$ function at $\Delta V=0.5$, 1.0, 1.5, 2.0, 2.5, and 3.0 \kms, with a fixed temperature of $T_{\rm rot}=20$ K. {\bf (c)} The relation between the $(1,1)$ optical depth $\tau_0$ and $R_{\rm sm}$ estimated at $T_{\rm rot}=20$ K and $\tau_{11}=2.0$. The slope of $\tau_0$-$R_{\rm sm}$ relation is almost not effected by $T_{\rm rot}$. }
\label{fig:cf_rsm}
\end{figure*}

Figure \ref{fig:cf_rsm}a shows $C_f(R_{\rm sm})$ sampled at temperatures from $T_{\rm rot}=10$ to 70 K. It shows that each $C_f(R_{\rm sm})$ relation has a clear and smooth variation trend with $R_{\rm sm}$. The slope of the $C_f(R_{\rm sm})$ relation continuously varies with $T_{\rm rot}$. 


Figure \ref{fig:cf_rsm}b shows the $C_f(R_{\rm sm})$ relations at line widths from $\Delta V=0.3$ to 3.0 \kms. As shown in Figure~\ref{fig:cf_rsm}b, the slope of the $C_f(R_{\rm sm})$ relation increases from $\overline{C_f/R_{\rm sm}}=0.25$ to 0.5 over the $\Delta V$ range. The changing of $C_f(R_{\rm sm})$ curves with $\Delta V$ should also reflect the changing of blending condition of the internal components in each HFC group as shown in Figure \ref{fig:spec_dv}a. This effect is now included in \emph{HFGR}. 


Figure \ref{fig:cf_rsm}a and \ref{fig:cf_rsm}b also show that all the $C_f(R_{\rm sm})$ curves are exactly converged at $(R_{\rm sm},C_f)=(1.0,0.9524)$, which represents the line intensity ratio at extremely low optical depth. When $\tau_0$ is very small, the HFC group intensities would become independent of $\Delta V$ and $T_{\rm rot}$. 

Since $R_{\rm sm}$ is a correction factor for optical depth $\tau_0$, we examined the relation between $R_{\rm sm}$ and $\tau_0$. For each (1,1) model spectrum, $\tau_0$ can be estimated from Equation (7). As shown in Figure  \ref{fig:cf_rsm}c, the two quantities are found to have a nearly linear relation of $\tau_0(1,1)=3.52(R_{\rm sm}-1)$. Since $\tau_0$ can be uniquely determined by $R_{\rm sm}$, it is not necessary to be independently considered in our calculation. 

There is still a caveat in using $R_{\rm sm}$ to estimate the optical depth due to the hyperfine intensity anomaly (HIA) \citep[][and references therein]{camarata15}, which would cause increased intensities of the F=1-2 (left isg) and 0-1 (right osg) components due to the over population at $F$=0 and $F$=1 states during the $J_K$=$2_1-1_1$ transition. In LTE condition, the HIA would only enhance the hyperfine components, but would not change the overall population of the (2,2) and (1,1) levels. In order to circumvent the HIA-effect to the $T_{\rm rot}$ calculation, one can consider to use two times of the F=1-0 (left osg) and F=2-1 (right isg) intensities to estimate the numerator in Equation (14), namely $\int T_{\rm mb}^{\rm isg+osg} {\rm d}\nu = 2 \int (T_{\rm mb}^{F=1-0}+T_{\rm mb}^{F=2-1}) {\rm d}\nu$. 

\subsection{Coefficients in the Polynomial Expression of $T_{\rm rot}$}\label{sec:hfgr_use}
At any $\Delta V$ and $T_{\rm rot}$ values, the $C_f(R_{\rm sm})$ relation exhibits a linear increasing or decreasing trend with a slight curvature. We thus tried to fit it using a two-order polynomial, 
\begin{equation}  
C_f=a_0+a_1(R_{\rm sm}-R_{\rm sm0})+a_2(R_{\rm sm}-R_{\rm sm0})^2,
\end{equation}
wherein $R_{\rm sm0}=1.0$, $a_0=0.9524$ represent the values at the convergent point (Figure \ref{fig:cf_rsm}a). The coefficients $a_{\rm 1,2}$ are constants for any individual $C_f(R_{\rm sm})$ relation, but would depend on $T_{\rm rot}$ and $\Delta V$. 

\begin{figure*}
\centering
\includegraphics[angle=0,width=0.9\textwidth]{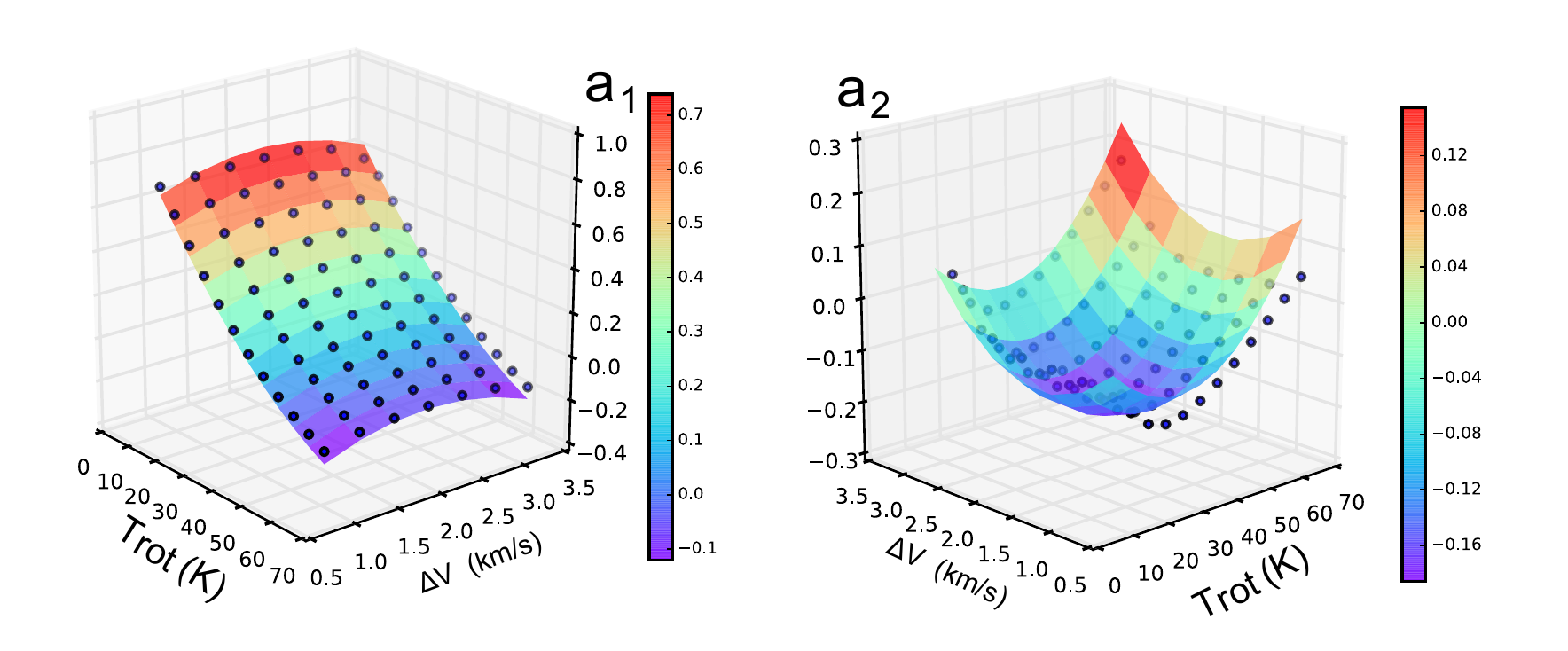}       \\     
\caption{\small The variation of $a_i~(i=1,2)$ as a function of $\Delta V$ and $T_{\rm rot}$. The dots indicate the $a_i$ values sampled from the modeled spectra. The surface in each panel represents best-fit 2D polynomial function as shown in Equation (16).}
\label{fig:ai_tv}
\end{figure*}

In Figure \ref{fig:ai_tv}, the solid dots represent $a_1$ and $a_2$ values sampled over the parameter space of $\Delta V$ and $T_{\rm rot}$. The numerical functions of $a_{1,2}(T_{\rm rot},\Delta V)$ are also fitted by two-order polynomials, 
\begin{equation}  
a_i=h_0+h_1\Delta V+h_2 T_{\rm rot}+h_3 \Delta V^2+ h_4 T_{\rm rot}^2,
\end{equation}
wherein coefficients $\{h_i\}$ are permanent constants that no longer depend on the parameters of $\tau_0$, $T_{\rm rot}$, or $\Delta V$. Based on the numerically sampled $a_i$, we found that $T_{\rm rot}$ and $\Delta V$ are independent in determining $a_i$. It is thus not necessary to have a crossing term of $\Delta V T_{\rm rot}$ in Equation (16).

In Figure \ref{fig:ai_tv}, the best-fit equations of $a_{i}(\Delta V, T_{\rm rot})$ are presented in false-color surfaces in each panel. We see that the surface of $a_i(\Delta V, T_{\rm rot})$ functions with the best-fit $\{h_i\}$ values are coincident with the sampled data points, suggesting that Equation (16) can closely describe the $a_i$ variation as a function of $T_{\rm rot}$ and $\Delta V$. 

\subsection{How to Perform the Recipe}\label{sec:hfgr_use}
\begin{figure*}
\centering
\includegraphics[angle=0,width=0.6\textwidth]{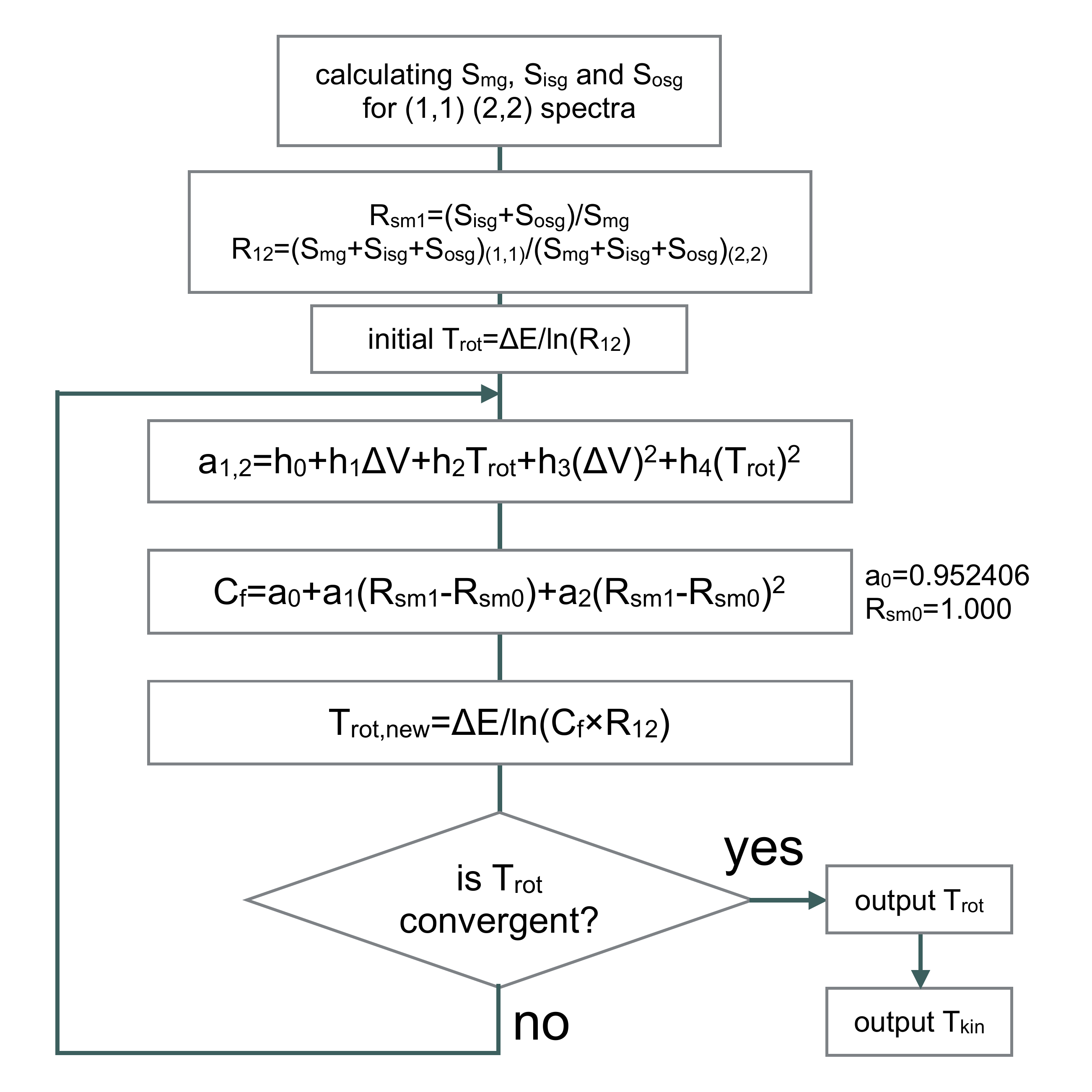}     \\     
\caption{\small The flow chart showing the calculation process of the recipe. S$_{\rm mg}$, S$_{\rm isg}$ and S$_{\rm osg}$ are integrated intensities of main group (mg), inner satellite group (isg) and outer satellite group (osg). $C_f$ is a correction factor for optical depth. The coefficients ${h_i}$ are constants and no longer depend on physical parameters $\tau_0$, $T_{\rm rot}$, or $\Delta V$. }
\label{fig:flow_chart}
\end{figure*}

The major steps of using the recipe are presented in a flow chart in Figure \ref{fig:flow_chart}. In calculation, an initial value of $T_{\rm rot}$ should be provided. It can be calculated from Equation (11) assuming $C_f=1.0$. And the line width $\Delta V$ can be measured from the (1,1) major group. As described above, compared to $T_{\rm rot}$, $\Delta V$ has a minor influence to $C_f$. We only need to ensure that $\Delta V$ is not largely deviated from the actual value so that its influence to $C_f$ can be corrected. $a_{1,2}$ and $C_f$ are then derived using Equation (16) and (15), respectively. And $T_{\rm rot}$ is calculated again using Equation (13). The calculation can be usually converged after several iterations. 

One can also estimate the kinetic temperature $T_{\rm kin}$. The recommended formula is from \citet{tafalla04},    
\begin{equation}    
T_{\rm kin}=\frac{T_{\rm rot}}{1-\frac{T_{\rm rot}}{\Delta E/k}\ln[{1+1.1\exp(-\frac{16}{T_{\rm rot}})]}},
\end{equation}
which is obtained from a Monte Carlo sampling of the $(J,K)$-level population as a function of $T_{\rm kin}$. 

\subsection{Accuracy of the recipe}\label{sec:hfgr_accuracy}
\begin{figure*}
\centering
\includegraphics[angle=0,width=0.9\textwidth]{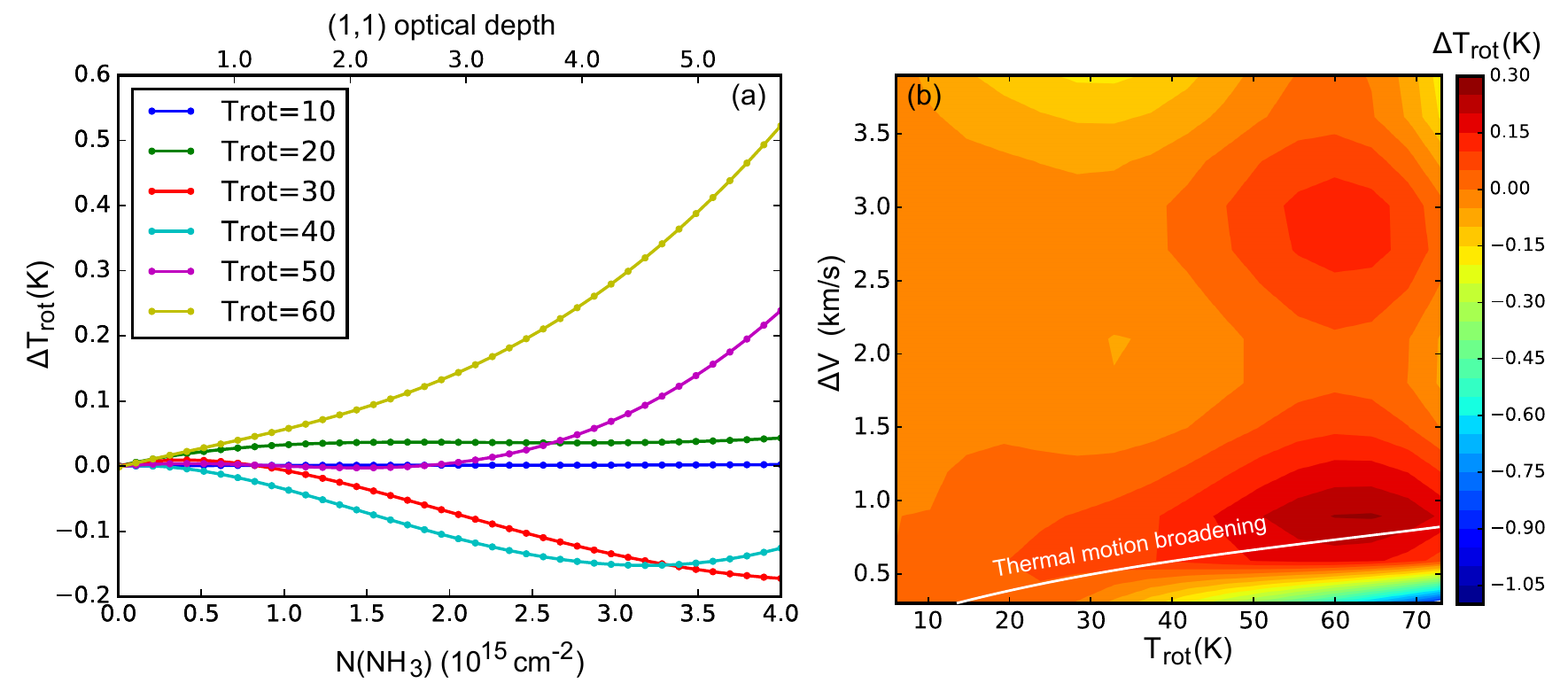}      \\     
\caption{\small Intrinsic $T_{\rm rot}$ error in our recipe. {\bf (a)} The deviation of $T_{\rm rot}$ from the actual value as a function of $\nht$ column density at different $T_{\rm rot}$ values. {\bf (b)} The standard deviation of $T_{\rm rot}$ in the parameter space of ($T_{\rm rot}$, $\Delta V$). In calculation, the optical depth of $(1_1^--1_1^+)$ is adopted to be a constant of $\tau_0 (1,1)=3.0$. The white curve denotes the line width due to the thermal motion under $T_{\rm K}$ estimated from Equation (17). The region bellow the curve would not exist in real condition. }
\label{fig:dt_ncol}
\end{figure*}

In order to test the accuracy of \emph{HFGR}, it is applied to a series of model spectra. The $(1_1^--1_1^+)$ and $(2_2^--2_2^+)$ model spectra are generated using Equation (3), (4), and (5). The input parameters include $T_{\rm rot}$, $\Delta V$, and $\tau_0$. Figure \ref{fig:dt_ncol}a shows the output $T_{\rm rot}$ deviation as a function of $N(\nht)$ (or $\tau_0$) at a number of $T_{\rm rot}$ values. The $T_{\rm rot}$ deviation turns out to increase moderately with $N(\nht)$ and remains $\Delta T_{\rm rot}<1.0$ K throughout the temperature range. In the high-mass dense molecular cores, the column densities have an average level of $N(\nht)=5\times10^{14}$ cm$^{-2}$ and only occasionally exceed $10^{15}$ cm$^{-2}$ \citep[e.g.][Table 6 therein]{lu14}. The $N(\nht)$ values in real cases are thus well covered in our calculation range, and the $T_{\rm rot}$ deviation due to the $N(\nht)$ would not be significant.   

Figure~\ref{fig:dt_ncol}b shows the $T_{\rm rot}$ deviation over the $(T_{\rm rot},\Delta V)$ parameter space at $\tau_0=3.0$. The $T_{\rm rot}$-error is quite small in the major fraction of the parameter space, and the $T_{\rm rot}$-error increases to a noticeable level only when $\Delta V$ is very small and $T_{\rm rot}$ is very high. There is an additional physical constraint that $\Delta V$ should be higher than the level of thermal velocity dispersion, that is $\Delta V_{\rm th} = (8 \ln 2~ k T_{\rm k}/\mu m_{\rm NH_3})^{1/2}$. This relation, as plotted in Figure \ref{fig:dt_ncol}b, represents a lower limit for the available $\Delta V$. The region bellow this curve would not exist in reality. Above this curve, the $T_{\rm rot}$ deviation is lower than 0.15 K and only slightly increases to $\Delta T\simeq 0.5$ K over the $T_{\rm rot}$ range. 

\subsection{The Modification at Low Temperatures}
\label{sec:hfgr_low_trot}
\begin{figure*}
\centering
\includegraphics[angle=0,width=0.4\textwidth]{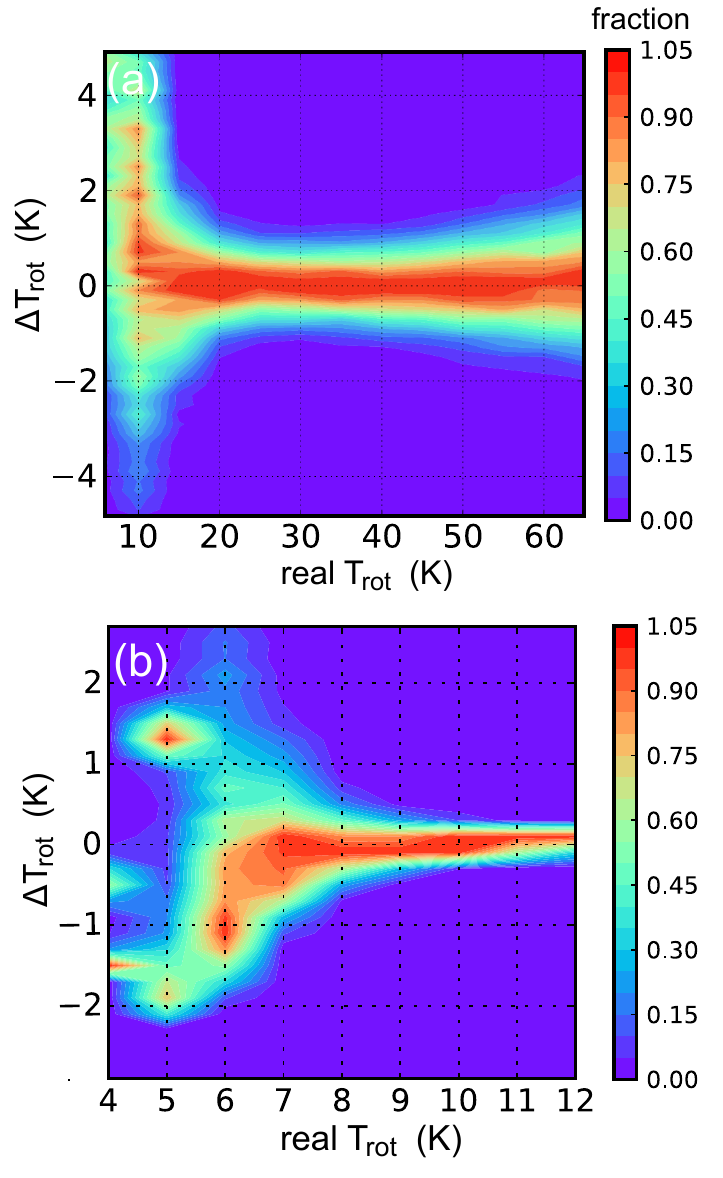}      \\      
\caption{\small {\bf (a)} The $T_{\rm rot}$-error distribution as a function of the actual $T_{\rm rot}$ for the current recipe. The temperature range is investigate with a step of 2.0 K. {\bf (b)} Same as (a), but for lower $T_{\rm rot}$ range and in calculation, each hyperfine group is fitted with a Gaussian profile to measure its integrated intensity. }
\label{fig:dt_trot}
\end{figure*}

At low temperatures, the (2,2) line will become very weak so that the $T_{\rm rot}$ uncertainty would largely increase. This problem exists in all three methods. Figure \ref{fig:dt_trot} shows the output $\Delta T_{\rm rot}$ distribution as a function of the real $T_{\rm rot}$ in the low temperature range for \emph{HFGR}. There is an drastic increase of $\Delta T_{\rm rot}$ dramatically increases when $T_{\rm rot}<15$ K. This is mainly due to the (2,2) emission becoming very weak at low temperatures. This uncertainty can be reduced by using a gaussian fitting to measure the HFC-group intensities. This can significantly eliminate the $T_{\rm rot}$ uncertainty as shown in Figure \ref{fig:dt_trot}b.

\subsection{Comparison between the \emph{HFGR} and other Methods}\label{sec:compare} 
We compared \emph{HFGR} with other two classical methods (Section \ref{sec:bg}) for the efficiency and accuracy. For \emph{Intensity ratio}, it seems that many studies used the peak line intensity of each HFC group \citep[e.g.][]{friesen09,ragan11,chira13,dhabal19}. Actually, as the second option, one can also use the integrated intensity of each HFC group instead of its peak value \citep[e.g.][]{williams18}. This would increase the signal-to-noise ratio (SNR). For example, if the line emission in a group extends over $N$ channels, the SNR of the integrated emission would be increased by a factor of $\sqrt{N}$. It is worthwhile to have a comparison between the two options, namely using the peak value or integrated emission of each HFC group.  

\emph{HFGR} and other two methods are applied to the model spectra to make comparison of their accuracies. We first investigate the $T_{\rm rot}$ variation at a constant SNR and $\sigma_v$ the three methods, wherein the input parameters are $T_{\rm rot}=20$ K, $\Delta V=1.0$ \kms, $\tau_0(1,1)=1.5$, and a spectral noise level of rms=0.2 K. The rms level corresponds to SNR$\simeq 20$ for the (1,1)-isg group. In each of the 2000 samplings, the rms noise is independently generated and added into the model spectra. In each sampling, $T_{\rm rot}$ is calculated from the noisy spectra using the three methods. 

\begin{figure*}
\centering
\includegraphics[angle=0, width=0.9\textwidth]{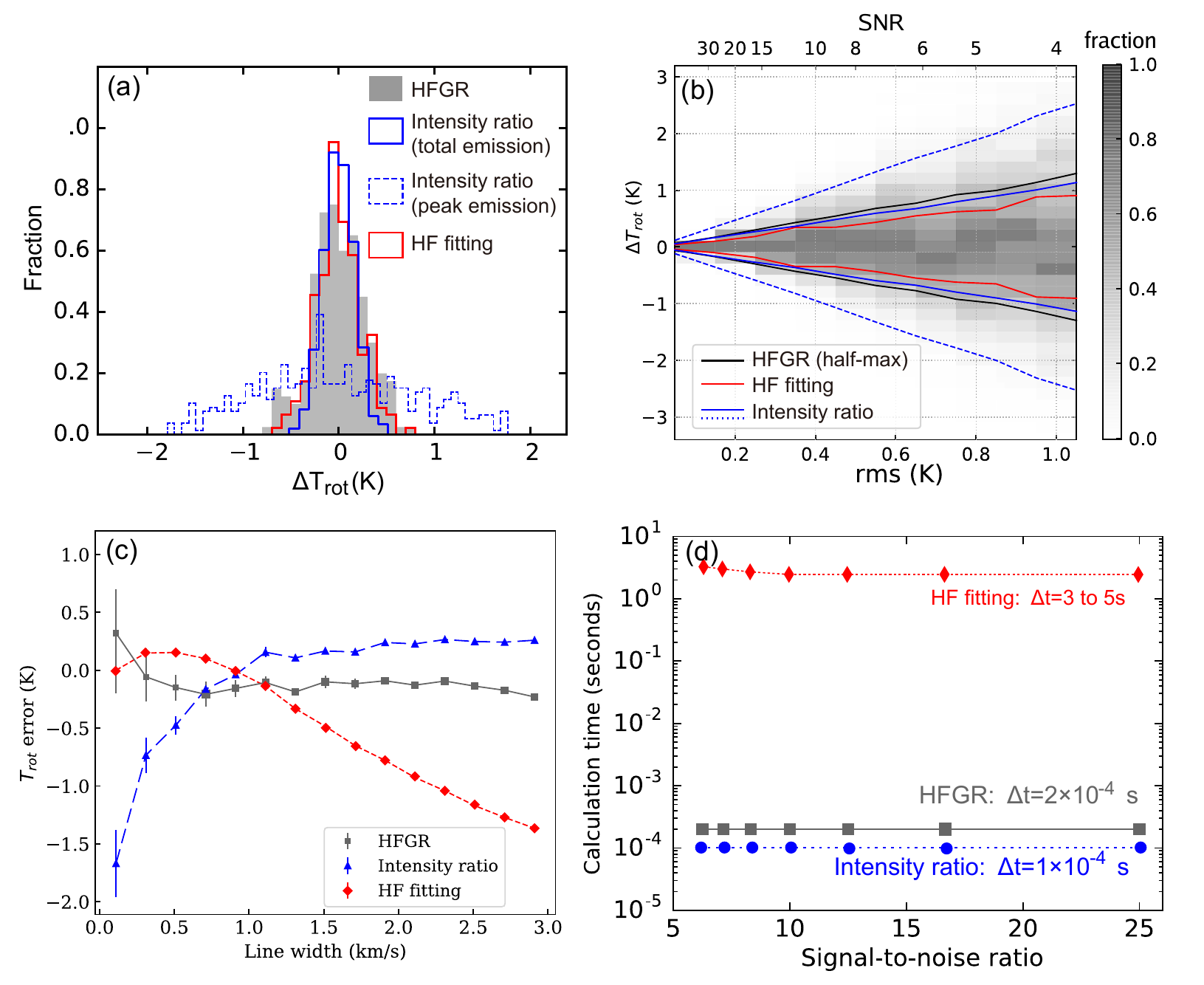}    \\  
\caption{Comparison of the three procedures of deriving $T_{\rm rot}$, wherein the real value is fixed at $T_{\rm rot}=20$ K. {\bf (a)} The $T_{\rm rot}$ error distribution in 2000 times of calculations for each procedure: \emph{Intensity ratio} (thin line), \emph{HF fitting}, and \emph{HFGR}. {\bf (b)} The $T_{\rm rot}$-error distribution as a function of rms noise level. At each rms value, the $\Delta T_{\rm rot}$ distribution is also obtained from 2000 samplings. The rms range is investigated with a step of 0.2 K. {\bf (c)} The calculation efficiency for each method as a function of the signal-to-noise ratio. The calculation time is that spent in deriving $T_{\rm rot}$ from one pair of $(1_1^--1_1^+)$ and $(2_2^--2_2^+)$ spectra. All the calculations are performed in one 2.3 GHz Intel Core i7 CPU. }
\label{fig:method_compare}
\end{figure*}

Figure~\ref{fig:method_compare}a shows the temperature variation $\Delta T_{\rm rot}$ relative to the actual value of $T_{\rm rot}=20$ K in all the samplings, wherein \emph{HFGR} and the \emph{HF fitting} turn out to have comparable variation of $\Delta T_{\rm rot}=\pm 0.5$ K. For \emph{intensity ratio}, the two options are both considered, namely using (i) total emission, and (ii) the peak value for each HF group are both investigated. As a result, option (i) exhibits a comparable $\Delta T_{\rm rot}$ variation, while option (ii) has a lager variation of $\Delta T_{\rm rot}=\pm 1$ K. The larger uncertainty is within our expectation since the peak $T_{\rm mb}$ of each hyperfine group is sensitive to the rms noise. 

Figure~\ref{fig:method_compare}b shows the $\Delta T_{\rm rot}$ distribution as a function of the rms noise. As expected, in each method, $\Delta T_{\rm rot}$ shows an increasing trend with the rms level. \emph{HF fitting} has the lowest $\Delta T_{\rm rot}$ over the rms range, while \emph{HFGR} and \emph{Intensity ratio} [option (i)] have slightly higher $\Delta T_{\rm rot}$ than \emph{HF fitting}. For all three methods, the variation can maintain a reasonable level of $\Delta T_{\rm rot}\leq 2.0$ K if the SNR is not too low (SNR$>4$). In comparison, \emph{Intensity ratio} [option (ii)] has much larger uncertainty that increases to $\Delta T_{\rm rot}=\pm 2.5$ K towards high rms level (1.0 K). It indicates that the option (ii) would have large uncertainty if applied to very noisy spectra. Therefore, when using the \emph{Intensity ratio} method to derive $T_{\rm rot}$, one should first attempt to follow option (i).

Figure~\ref{fig:method_compare}c shows the average $\Delta T_{\rm rot}$ as a function of the line width $\Delta V$. A notable feature is that $\Delta T_{\rm rot}$ becomes evidently deviated from the zero level over the $\Delta V$ range, in particular for \emph{HF fitting} and \emph{Intensity ratio}. And the three methods exhibit quite different variation trends. For \emph{HF fitting}, the value decreases to $\Delta T_{\rm rot}=-1.5$ K around $\Delta V=2.5$ \kms. The $\Delta T_{\rm rot}$ deviation with $\Delta V$ should be mainly due to the change of HFC blending conditions. At small $\Delta V$, the internal HFCs within each group can be sufficiently resolved by the hyperfine fitting. While $\Delta V$ becomes higher, the HFCs would gradually become blended, letting the fitted spectra be less constrained. 

\emph{Intensity ratio}, in contrast, exhibits a nearly opposite trend. It has relatively large deviation of $\Delta T_{\rm rot}=-1.7$ K at lower line width of $\Delta V=0.1$ to 0.5 \kms, and becomes converged to $\Delta T_{\rm rot}=\pm0.5$ K at $\Delta V>0.5$ \kms. The $\Delta T_{\rm rot}$ deviation at low $\Delta V$ is similar as that shown in Figure \ref{fig:spec_dv}. 

{\color{black} \emph{HFGR} exhibits an overall small variation in the $\Delta V$ range. This is within our expectation since the effect of $\Delta V$ is already considered in its calculation (Figure\,\ref{fig:cf_rsm}b).} Its $\Delta T_{\rm rot}$ slightly increases with a scale of 0.4 K only towards small $\Delta V$ ($\leq0.2$ \kms). At larger values of $\Delta V>0.5$ \kms, $\Delta T_{\rm rot}$ stays in a narrow range of $-0.2$ to 0 K. 

Figure~\ref{fig:method_compare}d shows the calculation time $\Delta t_{\rm cal}$ in the three methods. For each method, $\Delta t_{\rm cal}$ is measured from an average of 100 times of calculations run in the same computer. The \emph{HFGR} and \emph{Intensity ratio} has comparable $\Delta t_{\rm cal}=1$ to $2\times 10^{-4}$ seconds. \emph{HFGR} has slightly higher $\Delta t_{\rm cal}$ than \emph{Intensity ratio} because the \emph{HFGR} performs several iterations to optimize $C_f$ (Figure \ref{fig:flow_chart}). \emph{HF fitting} is much more time-consuming, with $\Delta t_{\rm cal}=3-4$ seconds, which is longer than the two other methods for a factor of $10^4$. This is because the \emph{HF fitting} would traverse over a large parameter space to look for the optimized values. 

Figure~\ref{fig:method_compare}d also shows a feature that for \emph{HF fitting}, its $\Delta t_{\rm cal}$ slightly increases towards the lowest SNR, while for \emph{Intensity ratio} and \emph{HFGR}, $\Delta t_{\rm cal}$ appears to be constant throughout the SNR range. As an explanation, for \emph{Intensity ratio} and \emph{HFGR}, its $\Delta t_{\rm cal}$ simply represents the time to run two or three analytical equations, thus would be clearly independent of the spectral noise. While \emph{HF fitting} would be affected by spectral shape, thus would be slightly delayed if the line profiles are largely disturbed by the noise. 

{\color{black} As a short summary, in the comparison test, \emph{HFGR} shows a balanced advantage between efficiency and accuracy, and has a relatively stable performance over a broad range of $\Delta V$ and SNR.}        

\subsection{Application to the Real Observational Data} 
\label{sec:obs_data} 
We tested the performance of \emph{HFGR} by applying it to the real observational data in Orion A region. This region contains compact and quiescent filament structures with moderate protostellar heating. The region is covered by the $\nht$ $(1_1^--1_1^+)$ and $(2_2^--2_2^+)$ observations \citep{li13,friesen17} with a sensitivity of $\sim0.1$ K, allowing us to make a detailed comparison of the three methods.

\begin{figure*}
\centering
\includegraphics[angle=0, width=0.9\textwidth]{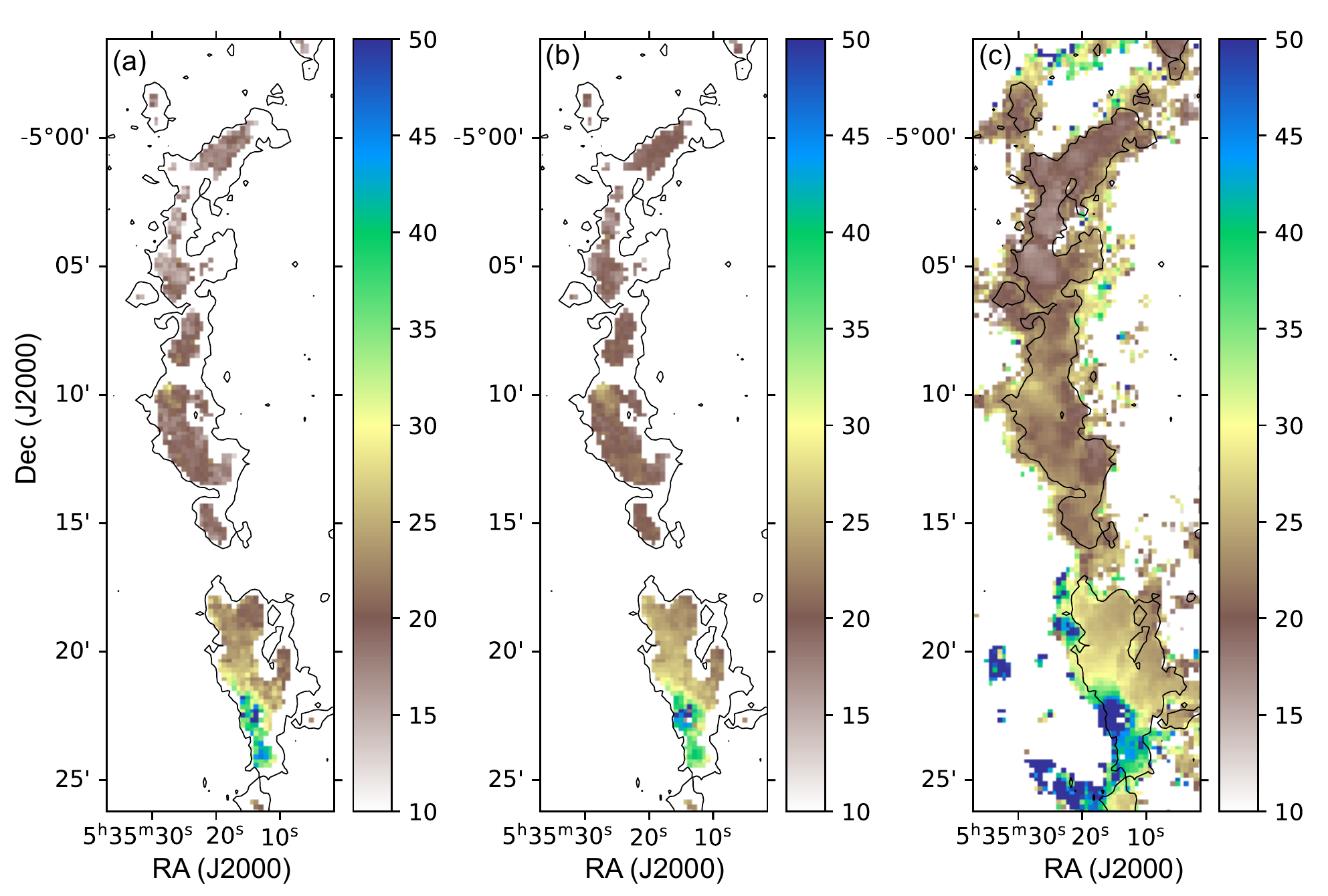} \\ 
\caption{\small The $T_{\rm rot}$ maps over Orion A North region calculated using three methods: {\bf (a)} \emph{HFGR}, {\bf (b)} \emph{Intensity Ratio}, and {\bf (c)} \emph{HF fitting}. The $T_{\rm rot}$ maps are calculated from the $\nht$ $(1_1^--1_1^+)$ and $(2_2^--2_2^+)$ data cubes observed with GBT \citep{friesen17}, which have a sensitivity of rms=0.1 K, velocity resolution of 0.07 \kms, and spatial resolution (beam size) of $32''$.}
\label{fig:trot_map}
\end{figure*}

Figure \ref{fig:trot_map}a to \ref{fig:trot_map}c show the $\nht$ $T_{\rm rot}$ maps of the three methods over the region of Orion A North. The three methods exhibit quite similar $T_{\rm rot}$ distributions over the $\nht$ emission region. In particular, around the Orion KL area, temperature sharply increase to $T_{\rm rot}\sim 45$ K, which is demonstrated by all three $T_{\rm rot}$ maps.

\begin{figure*}
\centering
\includegraphics[angle=0,width=0.8\textwidth]{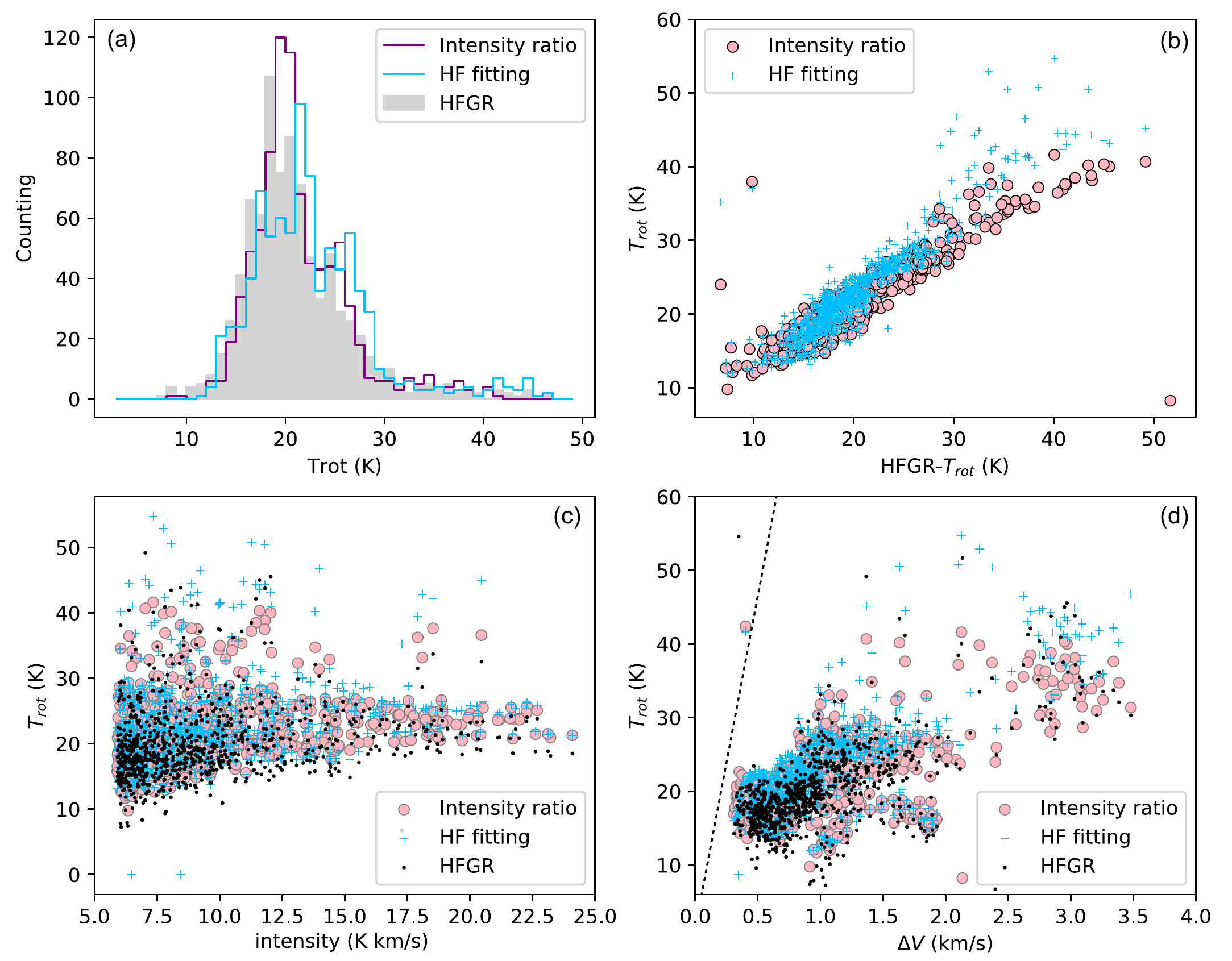} \\ 
\caption{{\bf (a)} The statistical distribution of the $T_{\rm rot}$ values for the pixels over the emission region above $5\sigma$ detection limit of the $\nht$ $(1_1^--1_1^+)$ image. {\bf (b)} Comparison of $T_{\rm rot}$ values derived from \emph{HFGR} and the other two methods. {\bf (c)} Relation between $T_{\rm rot}$ and $\nht$ $(1_1^--1_1^+)$ intensity for the three methods. {\bf (d)} Relation between $T_{\rm rot}$ and $\Delta V$ for the three methods. The dashed line represents the expected $\nht$ line width due to the thermal motion as a function of $T_{\rm rot}$, assuming $T_{\rm rot}$ and $T_{\rm kin}$ satisfying Equation (17). }  
\label{fig:trot_stat}
\end{figure*}

The statistical distribution of $T_{\rm rot}$ for the three methods are shown in Figure \ref{fig:trot_stat}a. For each method, the major fraction of the data points have quite similar and overlapped distribution profiles, concentrated in range of $T_{\rm rot}=10-30$ K, and a maximum distribution around $T_{\rm rot}=19$ K. The similar distributions suggests that the three methods have comparable accuracy in deriving $T_{\rm rot}$. 

The $T_{\rm rot}$ distributions of the different methods are compared from their $T_{\rm rot}$ values at the pixels within the $\nht$ emission region, as shown in Figure \ref{fig:trot_stat}b. The \emph{intensity ratio} and \emph{HF fitting} both exhibit a linear increasing trend with \emph{HFGR} over the range of $T_{\rm rot}=5$ to 50 K, with a variation of $\Delta T_{\rm rot}=\pm 3$ K. This also suggests that the three methods have not only similar $T_{\rm rot}$ ranges, but also spatially coherent $T_{\rm rot}$ variations. The only difference is that \emph{HF fitting} has a small fraction of the pixels at higher $T_{\rm rot}$ values than in other two methods. These pixels are located in the relatively high temperature range from 30 to 50 K. Bellow this range ($T_{\rm rot}<30$ K), the data of three methods are almost fully overlapped. In Figure \ref{fig:trot_map}, the high-temperature areas are mainly from Orion KL region, where the $\nht$ emission turns out to be weaker than the Northern part of the cloud. This could be due to the disruption of the $\nht$ by the radiation from the massive stars in M42 (Trapezium Cluster). As the consequence, the weakness of the line emission, in particular for the $(2_2^--2_2^+)$ line, would be responsible for the $T_{\rm rot}$ deviation in \emph{HF fitting}. In comparison, \emph{HFGR} and \emph{intensity ratio} have closely correlated $T_{\rm rot}$ distributions probably because they both use integrated intensities of the HFC groups. 

Figure \ref{fig:trot_stat}c shows the distribution of $T_{\rm rot}$ and the $(1_1^--1_1^+)$ integrated intensity. In this diagram, the three methods also exhibit largely overlapped distributions. It more clearly shows that the majority of high temperature points ($T_{\rm rot}>30$ K) have weak intensities, with $\int T_{\rm mb,11} {\rm d}V<12$ K \kms, which is also suggestive of the $\nht$ disruption in the hot region.    

Figure \ref{fig:trot_stat}d shows the distributions $T_{\rm rot}$ and the $(1_1^--1_1^+)$ line width $\Delta V_{11}$. The data points for the three methods are also largely overlapped. And they all exhibit a trend of increasing $T_{\rm rot}$ with $\Delta V_{11}$. The velocity due to the thermal motion can be calculated from $\sigma_{\rm th}=\Delta V_{\rm th}/\sqrt{8 \ln 2}=\sqrt{k_B T_{\rm kin}/m_{\rm \nht}}$. Assuming that $T_{\rm rot}$ and $T_{\rm kin}$ follow Equation (17), we estimated the theoretical relation of $\Delta V_{\rm th}(T_{\rm rot})$, as plotted in dashed line in Figure \ref{fig:trot_stat}d. The observed $\Delta V$ is shown to be much higher than $\Delta V_{\rm th}$ throughout the $T_{\rm kin}$ range, suggesting that the gas motion traced by $\nht$ is dominated by the non-thermal motion. Although the observed positive $\Delta V$-$T_{\rm rot}$ relation is coherent with the normal case that thermal motion increases with the temperature, it actually reflect another property that the more turbulent gas components have higher temperatures. This could also be largely contributed by Orion KL region where the intense radiation field and gas expansion are providing strong heating and dynamical perturbation to increase the turbulence and temperature at the same time. In general, the observational test shows that the data points of the three methods are largely overlapped in the broad $\Delta V$ range, suggesting that \emph{HFGR} should have a stable performance in the conditions of both low and high turbulence. 

\section{Summary}\label{sec:summary}
In order to improve the accuracy and efficiency in calculating the gas temperature using $\nht$, we constructed a new recipe of using the total intensities of hyperfine groups in $\nht$ $(1_1^--1_1^+)$ and $(2_2^--2_2^+)$ inversion lines to derive $T_{\rm rot}$. It is tested on the model spectra over a broad parameter space to guarantee the reliability. The python program for this calculation is provided freely\footnote{All the python codes are publicly available this web site: https://github.com/plotxyz/nh3\_trot.git}.

In building \emph{HFGR}, we made three major efforts to improve the $T_{\rm rot}$ calculation:

(1) \emph{HFGR} uses a group of empirical formulae (mainly Equation 13 to 16) based on forward radiative transfer calculations to derive $T_{\rm rot}$. The equations only rely on the intensities of HFC groups that requires no hyperfine fitting, thus the uncertainties due to the spectral profiles can be largely prevented. \emph{HFGR} can maintain an intrinsic uncertainty of $\Delta T_{\rm rot}<0.15$ K over a parameter space of $T_{\rm rot}=10$ to 70 K, $\Delta V=0.5$ to 3.5 \kms, and $N({\nht})<2\times10^{15}$ cm$^{-2}$. In comparison, the two other methods are both sensitive to the spectral shape, thus their accuracy could be significantly affected by $\Delta V$.

(2) Compared to \emph{hyperfine fitting}, \emph{HFGR} can substantially reduce a large amount of computational load because it does not require a traversal over the parameter space. 

(3) When applied to the noisy spectra, \emph{HFGR} can maintain an uncertainty at the level of $\Delta T_{\rm rot}\leq 1.0$ K (1 $\sigma$) when SNR$>4$. One can directly estimate the error from the relation between $\Delta T_{\rm rot}$ and the spectral noise.

\emph{HFGR} is applied to the $\nht$ lines observed in Orion A North region. As a result, the derived $T_{\rm rot}$ map exhibits a comparable result with \emph{HF fitting} and \emph{Intensity Ratio}. It suggests that \emph{HFGR} can have an unbiased temperature measurement from the observational data.

\section{Acknowledgement}
We thank the referee for very detailed comments that help improve the scientific analysis. This work is supported by the National Natural Science Foundation of China No. 11988101, No. 11725313, No. 11403041, No. 11373038, No. 11373045, CAS International Partnership Program No. 114A11KYSB20160008, the China Scholarship Council No. 201704910686, and the Young Researcher Grant of National Astronomical Observatories, Chinese Academy of Sciences.

\section{Data Availability}
The data underlying this article will be shared on reasonable request to the corresponding author.


\bsp 
\setcounter{table}{0}

\begin{landscape}
\begin{table}
\caption{Hyperfine intensities of the $\nht$ $(1_1^--1_1^+)$ and $(2_2^--2_2^+)$.}
\label{tbl:hf_pars}
\begin{tabular}{cccccccccccccc}
\hline\hline
Hyper-fine      &  HFC       &  $F' \to F$        &  $F'_1 \to F_1$     &  Frequency     &  Relative            &  Velocity&  HFC &  $F' \to F$     &  $F'_1 \to F_1$   &  Frequency  & Relative &   Velocity     \\
Group           &  number    &  \quad             &  \quad              &  (kHz)         &  Intensities$^a$     &  \kms        &  number           &  \quad          &  \quad            &  (kHz)      &  Intensities$^a$  &  \kms         \\
\hline                                                                                 
&$\nht~(1_1^--1_1^+)$    &  \quad             &  \quad              &  \quad         &  \quad               &  \quad       &  $\nht~(2_2^--2_2^+)$     &  \quad          &  \quad            &  \quad      &  \quad            &  \quad      \\
\hline                                                                                 
osg.1           &  1         &  $1/2,1/2$         &  (0,1)              &  -1568.49      &  $1/27$              &  -19.84      &  1               &  $3/2,3/2$      &  (1,2)            &  -2099.03   &  $1/300$     &  -26.53  \\
\quad           &  2         &  $1/2,3/2$         &  (0,1)              &  -1526.96      &  $2/27$              &  -19.32      &  2               &  $3/2,5/2$      &  (1,2)            &  -2058.26   &  $3/100$     &  -26.01  \\
\quad           &  \quad     &  \quad             &  \quad              &  \quad         &  \quad               &  \quad       &  3               &  $1/2,3/2$      &  (1,2)            &  -2053.46   &  $1/60$      &  -25.95  \\       
\hline                                                                                                                                                                                                               
isg.1           &  3         &  $3/2,1/2$         &  (2,1)              &  -623.31       &  $5/108$             &  -7.89       &  4               &  $7/2,5/2$      &  (3,2)            &  -1297.08   &  $4/135$     &  -16.39  \\
\quad           &  4         &  $5/2,3/2$         &  (2,1)              &  -590.34       &  $1/12$              &  -7.47       &  5               &  $5/2,3/2$      &  (3,2)            &  -1296.10   &  $14/675$    &  -16.38  \\
\quad           &  5         &  $3/2,3/2$         &  (2,1)              &  -580.92       &  $1/108$             &  -7.35       &  6               &  $5/2,5/2$      &  (3,2)            &  -1255.33   &  $1/675$     &  -15.86  \\
\hline
mg              &  6         &  $1/2,1/2$         &  (1,1)              &  -36.54        &  $1/54$              &  -0.46       &  7               &  $3/2,1/2$      &  (1,1)            &  -44.51     &  $1/60$      &  -0.56   \\
\quad           &  7         &  $3/2,1/2$         &  (1,1)              &  -25.54        &  $1/108$             &  -0.32       &  8               &  $5/2,3/2$      &  (2,2)            &  -41.81     &  $1/108$     &  -0.53   \\
\quad           &  8         &  $5/2,3/2$         &  (2,2)              &  -24.39        &  $1/60$              &  -0.31       &  9               &  $7/2,5/2$      &  (3,3)            &  -41.44     &  $8/945$     &  -0.52   \\
\quad           &  9         &  $3/2,3/2$         &  (2,2)              &  -14.98        &  $3/20$              &  -0.19       &  10              &  $5/2,5/2$      &  (2,2)            &  -1.05      &  $7/54$      &  -0.013  \\
\quad           &  10        &  $1/2,3/2$         &  (1,1)              &  5.85          &  $1/108$             &   0.07       &  11              &  $3/2,3/2$      &  (2,2)            &  -1.05      &  $1/12$      &  -0.013  \\
\quad           &  11        &  $5/2,5/2$         &  (2,2)              &  10.52         &  $7/30$              &   0.13       &  12              &  $7/2,7/2$      &  (3,3)            &   0.31      &  $8/35$      &  0.0039  \\
\quad           &  12        &  $3/2,3/2$         &  (1,1)              &  16.85         &  $5/108$             &   0.21       &  13              &  $5/2,5/2$      &  (3,3)            &   0.31      &  $32/189$    &  0.0039  \\
\quad           &  13        &  $3/2,5/2$         &  (2,2)              &  19.93         &  $1/60$              &   0.25       &  14              &  $3/2,3/2$      &  (1,1)            &   1.05      &  $1/12$      &  0.013   \\
\quad           &  \quad     &  \quad             &  \quad              &  \quad         &  \quad               &  \quad       &  15              &  $1/2,1/2$      &  (1,1)            &   1.05      &  $1/30$      &  0.013   \\
\quad           &  \quad     &  \quad             &  \quad              &  \quad         &  \quad               &  \quad       &  16              &  $3/2,5/2$      &  (2,2)            &   39.71     &  $1/108$     &  0.50    \\
\quad           &  \quad     &  \quad             &  \quad              &  \quad         &  \quad               &  \quad       &  17              &  $5/2,7/2$      &  (3,3)            &   42.04     &  $8/945$     &  0.53    \\
\quad           &  \quad     &  \quad             &  \quad              &  \quad         &  \quad               &  \quad       &  18              &  $1/2,3/2$      &  (1,1)            &   46.61     &  $1/60$      &  0.59    \\
\hline
isg.2           &  14        &  $1/2,3/2$         &  (1,2)              &  571.79        &  $5/108$             &   7.23       &  19              &  $5/2,5/2$      &  (2,3)            &  1254.58    &  $1/675$     &  15.85   \\
\quad           &  15        &  $3/2,3/2$         &  (1,2)              &  582.79        &  $1/108$             &   7.37       &  20              &  $3/2,5/2$      &  (2,3)            &  1295.34    &  $14/675$    &  16.37   \\
\quad           &  16        &  $3/2,5/2$         &  (1,2)              &  617.70        &  $1/12$              &   7.81       &  21              &  $5/2,7/2$      &  (2,3)            &  1296.33    &  $4/135$     &  16.38   \\
\hline                                                                                                                                                                                                                          
osg.2           &  17        &  $1/2,1/2$         &  (1,0)              &  1534.05       &  $1/27$              &  19.41       &  22              &  $3/2,1/2$      &  (2.1)            &  2053.46    &  $1/60$      &  25.95   \\
\quad           &  18        &  $3/2,1/2$         &  (1,0)              &  1545.05       &  $2/27$              &  19.55       &  23              &  $5/2,3/2$      &  (2.1)            &  2058.26    &  $3/100$     &  26.01   \\
\quad           &  \quad     &  \quad             &  \quad              &  \quad         &  \quad               &  \quad       &  24              &  $3/2,3/2$      &  (2.1)            &  2099.03    &  $1/300$     &  26.53   \\
\hline
\end{tabular} \\
$a.$ { The Hyperfine intensities are taken from \citet{kukolich67,poynter75,mangum15}. The sum of the relative intensities is 1.0.}
\end{table}
\end{landscape}


\begin{table}
\begin{center}
\caption{The Physical Parameters of The $\nht$ Transitions.}
\label{tbl:line_pars}
\begin{tabular}{llcccc}
\hline
Transition    & Frequency$^a$         &  $E_{\rm u}$    &  $A^b$                  & $\gamma^c$                       & $n_{\rm crit}^d$ \\
\quad         & (GHz)                 &   (K)           &  ($10^{-7}$s$^{-1}$)    & ($10^{-11}$cm$^3$s$^{-1}$)       & ($10^3$ cm$^{-3}$) \\
\hline
$(1_1^--1_1^+)$         & 23.69449              &  24.35          & 1.86,5.58               & 8.3,9.5                          & 2.0,6.7     \\
$(2_2^--2_2^+)$         & 23.72263              &  65.34          & 0.83,6.63               & 11,13                            & 0.6,6.0     \\
\hline
\end{tabular} \\
\end{center}
$a.${ $\nht$ $(1_1^--1_1^+)$ and $(2_2^--2_2^+)$ inversion transition frequencies given by \citep{kukolich67}. } \\
$b.${ Einstein $A$ coefficients were previously measured by \citet{osorio09,poynter75,mangum15}. The current values are from \citet{mangum15}. The two values correspond to the lower and upper limits for all the HFCs, respectively. } \\
$c.${ Collisional coefficients $\gamma$ are taken from \citet{danby88}. For each transition, the two values correspond to that at the temperature of $T_{\rm kin}=10$ and 100 K, respectively. } \\
$d.${ Critical density of each transition, the two values correspond to the lower and upper limits among all the hyperfine groups. } \\
\end{table}

\begin{figure*}
\centering
\includegraphics[angle=0,width=0.9\textwidth]{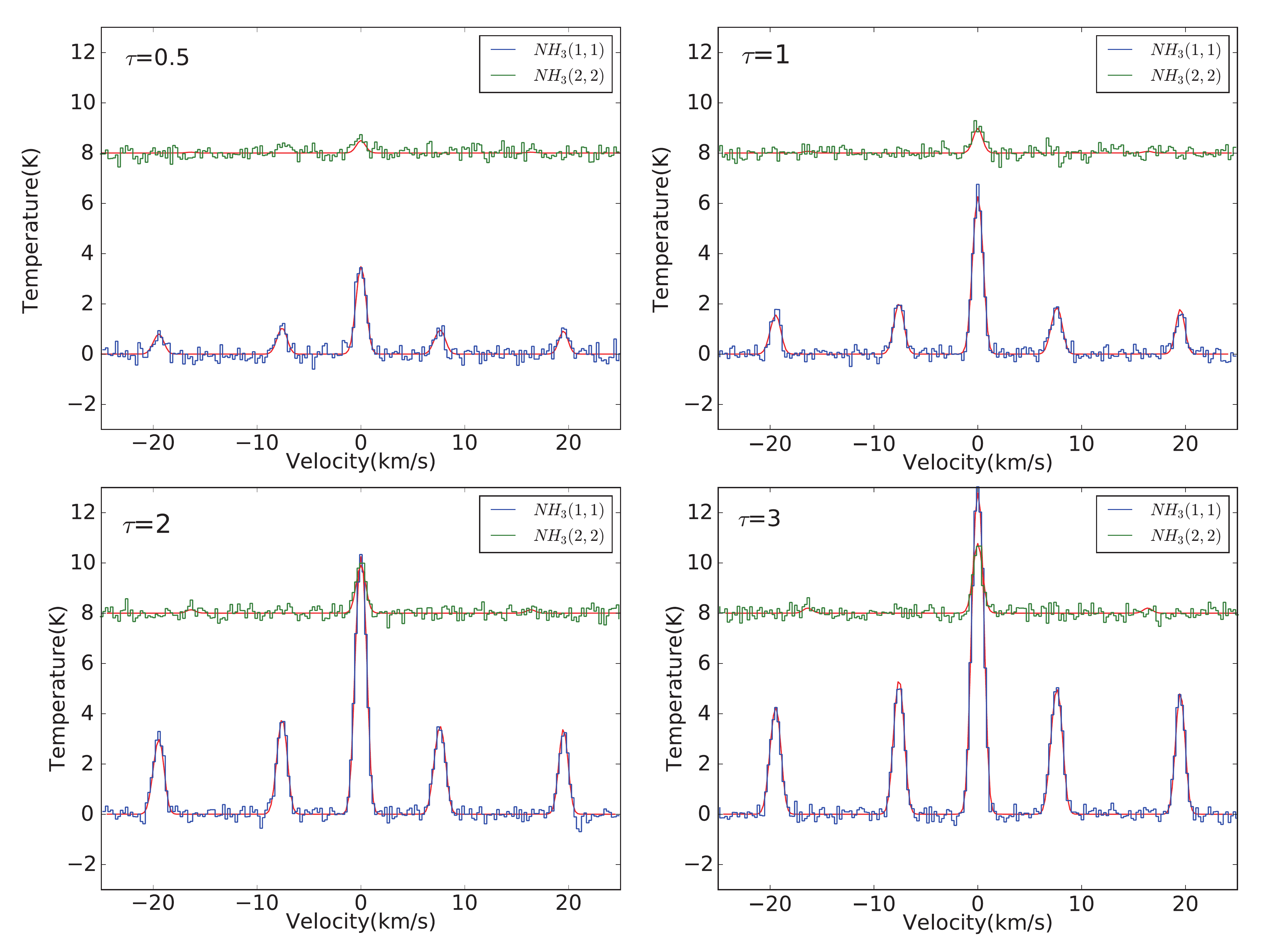} \\ 
\caption{\small The modeled spectra with a series of optical depths. The spectra are generated using the radiative transfer modeling as shown in Equation (1) to (3). The physical parameters are $T{\rm rot}=20$ K, $\Delta V=1.0$ \kms. In each spectrum the rms noise is set to be 0.3 K.}
\label{fig:ani4in1}
\end{figure*}

\label{lastpage}
\end{document}